\documentclass[a4paper,11pt]{article}
\pdfoutput=1 
\usepackage{jheppub} 
\usepackage[T1]{fontenc} 

\newcommand{\myroundedbrackets}[1]{\left(#1\right)}

\newcommand{\pq}[2]{$\{#1,#2\}$}

\usepackage{xcolor}

\usepackage{amsmath}
\usepackage{amsfonts}          
\usepackage{amssymb} 
\usepackage{mathrsfs}
\usepackage{braket}
\usepackage{physics}
\usepackage{amsthm}
\usepackage{stmaryrd}
\usepackage{cancel}
\usepackage{mathtools}
\usepackage{dsfont}
\usepackage{bbold}
\usepackage{csquotes}
\usepackage{setspace}
\usepackage{zref-savepos}
\usepackage{diagbox}
\usepackage{caption}
\usepackage{subcaption}
\usepackage{graphicx}
\usepackage{hyperref}
\usepackage[normalem]{ulem}

\def\avg#1{\left\langle#1\right\rangle}

\def\abs#1{\left|#1\right|}
\def\kc#1{\left(#1\right)}
\def\kd#1{\left[#1\right]}
\def\ke#1{\left\{#1\right\}}

\def\sgn{{\rm sgn}}

\def\tint{{\text{int}}}

\def\nn{\nonumber}
\def\pa{\partial}

\newcounter{NoTableEntry}
\renewcommand*{\theNoTableEntry}{NTE-\the\value{NoTableEntry}}

\usepackage{tikz}
\usetikzlibrary{decorations}
\usetikzlibrary{shapes.geometric}
\usetikzlibrary{calc}

\makeatletter
\DeclareRobustCommand{\iscircle}{\mathord{\mathpalette\is@circle\relax}}
\newcommand\is@circle[2]{%
  \begingroup
  \sbox\z@{\raisebox{\depth}{$\m@th#1\bigcirc$}}%
  \sbox\tw@{$#1\square$}%
  \resizebox{!}{\ht\tw@}{\usebox{\z@}}%
  \endgroup
}
\makeatother

\makeatletter
\newcommand{\doublewidetilde}[1]{{%
  \mathpalette\double@widetilde{#1}%
}}
\newcommand{\double@widetilde}[2]{%
  \sbox\z@{$\m@th#1\widetilde{#2}$}%
  \ht\z@=.9\ht\z@
  \widetilde{\box\z@}%
}
\makeatother

\makeatletter
\newcommand{\doubletilde}[1]{{%
  \mathpalette\double@tilde{#1}%
}}
\newcommand{\double@tilde}[2]{%
  \sbox\z@{$\m@th#1\tilde{#2}$}%
  \ht\z@=.9\ht\z@
  \tilde{\box\z@}%
}
\makeatother

\graphicspath{{Figures/}}

\def\kc#1{\left(#1\right)}
\def\kd#1{\left[#1\right]}
\def\ke#1{\left\{#1\right\}}

\title{Discrete JT gravity as an Ising model}

\author[a]{Johanna Erdmenger,}
\author[a]{Jonathan Karl,}
\author[a]{Yanick Thurn,}
\author[b]{Matthias Vojta}
\author[a]{and \\ Zhuo-Yu Xian}

\affiliation[a]{Institute for Theoretical Physics and Astrophysics and Würzburg-Dresden Cluster of Excellence ct.qmat, Julius-Maximilians-Universität Würzburg, Am Hubland, 97074 Würzburg, Germany}

\affiliation[b]{ Institute for Theoretical Physics and Würzburg-Dresden Cluster of Excellence ct.qmat,
 Technische Universität Dresden, 01062 Dresden, Germany}

\emailAdd{johanna.erdmenger@uni-wuerzburg.de}

\abstract{\noindent Inspired by the program of discrete holography, we show that Jackiw-Teitelboim (JT) gravity on a hyperbolic tiling of Euclidean AdS$_2$ gives rise to an Ising model on the dual lattice, subject to a topological constraint. The Ising model involves an asymptotic boundary condition with spins pointing opposite to the magnetic field. The topological constraint enforces a single domain wall between the spins of opposite direction, with the topology of a circle. The resolvent of JT gravity is related to the free energy of this Ising model, and the classical limit of JT gravity corresponds to the Ising low-temperature limit. We study this Ising model through a Monte Carlo approach and a mean-field approximation. For finite truncations of the infinite hyperbolic lattice,  the map between both theories is only valid in a regime in which the domain wall has a finite size. For the extremal cases of large positive or negative coupling, the domain wall either shrinks to zero or touches the boundary of the lattice. This behavior is confirmed by the mean-field analysis.
We expect that our results may be used as a starting point for establishing a holographic matrix model duality for discretized gravity.
}

\begin{document} 
\maketitle
\flushbottom

\section{Introduction}

The AdS/CFT correspondence \cite{Maldacena:1997,Witten:1998qj,Gubser:1998bc} is one of the most remarkable discoveries in theoretical physics in recent history. It fundamentally changed our perspective on gravity, while also providing tools to study strongly correlated systems. An important ingredient of this duality is the holographic principle \cite{tHooft:1993dmi,Susskind:1994vu}, which states that the information content of a gravitational theory in $d+1$-dimensions is stored on the $d$-dimensional boundary, where it is described by a quantum field theory without gravity. The AdS/CFT correspondence arises from string theory, as a duality between open and closed strings. It is therefore formulated in terms of continuous variables, and in particular, continuous spacetime. \\ The goal of the program known as \textit{discrete holography} \cite{Axenides:2013iwa,Axenides:2019lea,Axenides:2022zru,Asaduzzaman2020,Brower2021,Asaduzzaman:2021bcw,Brower2022,Gesteau:2022hss,Basteiro:2022pyp,Basteiro2022,Basteiro2022a,Yan:2018nco,Yan:2019quy,Petermann2023,Chen:2023cad,Dey:2024jno} is to generalize the holographic principle to discrete spacetimes. One motivation for this is to test how generally the holographic principle is valid, beyond the string theory context.  Moreover, discrete holography is motivated by the realization of hyperbolic lattices in experiments \cite{Kollar,Boettcher:2019xwl,RonnyTopolectric2018,dong2021topoelectric,lenggenhager2021electric}, as well as by the mathematical characterization of these lattices \cite{Magnus1974,CoxeterMoser,coxeter_1997,Gesteau:2022hss}. Based on the properties of the hyperbolic lattices, aperiodic spin chains have been proposed as candidates for  boundary theories \cite{Basteiro2022,Basteiro2022a}. Despite these results in the direction of discrete holography, a complete duality involving two precisely characterized theories, in analogy to continuum AdS/CFT, has not been realized at the dynamical level.\\
In this work we provide a further step toward establishing discrete holography by studying Jackiw-Teitelboim (JT) gravity \cite{TEITELBOIM198341,JACKIW1985343} on a two-dimensional hyperbolic lattice. Continuous JT gravity is a dilaton gravity theory with linear dilaton potential. It provides a solvable toy model for quantum gravity \cite{Mertens:2022irh}. This theory has been studied intensively in recent years in the holographic context \cite{Almheiri:2014cka,MaldacenaYang2016,Yang2018,Saad2019,Stanford2020,Iliesiu:2020zld,Iliesiu2021,Penington:2023dql,Kolchmeyer:2023gwa,Nogueira:2021ngh,Banerjee:2023eew}, together with its supersymmetric counterpart \cite{Stanford:2019vob,Okuyama:2020qpm,Turiaci:2023jfa} and more general dilaton gravity theories \cite{Maxfield2020,Witten:2020wvy,Collier2023}. In particular, it was proven that JT gravity is holographically dual to a matrix integral in the double scaling limit \cite{Saad2019}. The three-dimensional version of this duality, which involes random CFTs and tensor models, is considered in \cite{Belin:2023efa,Jafferis:2024jkb}.\\
It is thus natural to investigate the discrete analog of JT gravity as a bulk theory for discrete holography. In particular, we consider the disk partition function that corresponds to the leading order in the topological expansion of the path integral. Suppressing the higher genus contributions corresponds to the semiclassical approximation. As in the continuous case, the path integral on the disk topology is obtained from a sum over closed boundary paths, weighted by the enclosed area. We show that this sum is equivalent to a sum over spin configurations on the dual lattice, subject to a topological constraint. In particular, the resolvent function of discrete JT gravity is given by the partition function of an Ising model on the dual lattice. The resolvent is related to the disk partition function by a Laplace transform, with integration over the boundary length of the hyperbolic disk. Due to its gravity origin, the Ising model satisfies a topological condition that constrains the spins pointing up to form a single domain without holes. The spins outside of this domain have to point down. This constraint imposes a significant restriction on the space of allowed spin configurations and makes the model very different from the standard Ising model. In particular, it does not show anti-ferromagnetic behavior or geometric frustration \cite{jalagekar2023geometric}, even for infinitely negative coupling constants.\footnote{This is the reason why we do not refer to the coupling constant as ferromagnetic or anti-ferromagnetic, but denote it as positive or negative in the respective cases.}\\
Without this constraint, the interaction term of the hyperbolic Ising model was investigated  numerically in \cite{breuckmann2020critical}. Moreover, it was previously studied as a bulk model in discrete holography \cite{Asaduzzaman:2021bcw}. In that work, the Ising model is considered as a discrete analog of a quantum field theory living in the bulk, for example, a scalar field. This is different from the approach of the present paper, in which the Ising model arises as an effective description of the gravity theory. Furthermore, the hyperbolic Ising model has also appeared in the holographic computation of the second Renyi entropy through a tensor network \cite{Hayden:2016cfa}. There, the Ising spins are defined on the vertices of the network, and the second Renyi entropy is given by the difference in free energy between two generalized Ising models with different boundary conditions. \\
We numerically solve the Ising model with the topological constraint described above using a Monte Carlo approach. Since this necessarily includes a finite truncation of the infinite hyperbolic lattice, the relation to the original gravity theory is only valid in a certain regime. This regime is specified by the domain wall between spins pointing in opposite directions being located away from the boundary of the finite lattice. This is similar in spirit to recent work on the continuous theory, where JT gravity was studied at finite cutoff \cite{Stanford2020,Iliesiu:2020zld}. In \cite{Stanford2020}, the disk partition function is evaluated by studying a self-avoiding random walk on the hyperbolic plane. 
The self-avoiding condition imposes a disk topology. We find a related condition here: the domain of spins pointing up has to be simply connected. We further comment on the similarities and differences of our work with \cite{Stanford2020} in section \ref{Sec:Conclussion}.\\
We numerically evaluate the free energy of the Ising model. To this effect, we anneal the system, i.e.~we lower the temperature after a large number of update steps. We show how the parameters of the Ising model are related to the parameters of discrete JT gravity.  In our case, lowering the temperature of the Ising model corresponds to taking a semiclassical limit in the gravity theory. We use this approach to determine the ground state, which we find to be non-degenerate due to the topological constraint. Evaluating the ground state for different values of the coupling constant, we find three different regimes in the phase structure of the Ising model, with the following properties:\\
For small positive values of the coupling constant, in units of the magnetic field, we find a regime in which the spin domain has finite size. This corresponds to discrete JT gravity, as argued for the infinite lattice above, and the free energy as a function of the coupling constant is related to the resolvent of the gravity theory as a function of energy. The numerical analysis describes this regime well, as long as the coupling constant is smaller than the magnetic field, since then the numerical solution converges to the ground state within finite time. In this case, the numerical results for the Ising ground state match the behaviour expected from discrete JT gravity at finite cutoff, even though we derived the relation between JT gravity and Ising model for an infinite lattice. However, as soon as the coupling constant and the magnetic field are roughly of the same size, the numerical analysis of the intermediate regime breaks down:
for these parameter values, the time required by  the simulation to reach the ground state diverges, and it is never reached. We further comment on this phenomenon in the discussion of section \ref{sec:Ising model}. \\ 
In addition, by analytic continuation of the coupling constant, we find two additional extremal regimes, both numerically and using a mean-field approach. These regimes allow for an unambiguous determination of the ground state, while not having a gravity interpretation. The magnetically dominated regime includes negative, as well as small positive values of the coupling constant (in units of the magnetic field). In the magnetically dominated regime, the size of the domain of spins pointing up is maximized, and the free energy is a linear function of the coupling constant. We note that in this regime, the numerical analysis determines only the qualitative behavior of the free energy, due to finite size effects of the lattice. The coupling-dominated regime corresponds to a large positive coupling constant. Here the size of the spin-up domain is minimised, i.e.~it shrinks to zero size and all spins point down. The free energy approaches a constant as function of the coupling. \\ 
To validate the numerical findings, we also use a mean-field approach to study the Ising model on an infinite hyperbolic lattice. The mean magnetic field is considered as a local quantity that depends on the radial distance, i.e.~on the lattice layer.
We derive a self-consistency condition for the local magnetic field, which we subsequently solve. Depending on whether the coupling constant is larger or smaller than the external magnetic field, the ground state of the model corresponds to all spins pointing either upwards or downwards. The mean-field approach thus correctly describes the two asymptotic coupling constant regimes. The intermediate regime is not present in the mean-field approximation, since the assumption that the local magnetization only depends on the radial distance does not hold in this regime.\\
Our analytical results provide an explicit map between discrete JT gravity and an Ising model.  These are  confirmed numerically for a certain parameter regime even for a finite lattice. In this regime, the numerical approach thus provides a novel way of simulating two-dimensional gravity theories. \\
Our findings realize a dynamical discrete gravity theory, which provides a further step towards establishing the AdS/CFT correspondence for discrete spaces. To extend these results to a full discrete holographic duality, involving two dynamical theories both in the bulk and at the boundary, it remains to
determine the dual boundary theory. It is known that the continuous JT gravity theory is dual to a double-scaled matrix model \cite{Saad2019}. Our results provide crucial input towards establishing a similar duality in the discrete case. We provide a more detailed discussion of future directions in section \ref{Sec:Conclussion}.\\

The paper is organized as follows. In Section \ref{sec:Discrete JT}, we obtain the disk partition function of discrete JT gravity on a hyperbolic lattice by adapting the corresponding continuum calculation.  In Section \ref{sec:Ising model}, we show that discrete JT gravity is equivalent to an Ising model on the dual lattice, subject to a topological constraint. We study the free energy of the Ising model, which is related to the resolvent of the gravity theory. Section \ref{sec:mean field} provides an alternative approach towards the Ising model using mean-field theory, and it is based on the underlying inflation rules of a hyperbolic tiling. Finally, we conclude in Section \ref{Sec:Conclussion} with a discussion of the main results of this paper,
as well as reporting perspectives for future work. Some technical details are reported in the Appendix
\ref{apx:counterterm}.

\section{JT gravity on a hyperbolic lattice}
\label{sec:Discrete JT}

We begin by introducing JT gravity and review the calculation of the disk partition function. Afterwards, we explain the canonical discretization of the hyperbolic plane via so-called hyperbolic tilings. Finally, we define JT gravity on a hyperbolic lattice and repeat the calculation of the disk partition function in the discrete case. 

\subsection{Review of JT gravity on the hyperbolic plane}

\begin{figure}[t]
    \centering
    \begin{subfigure}[t]{0.45\textwidth}
        \centering
        \includegraphics[width=\linewidth]{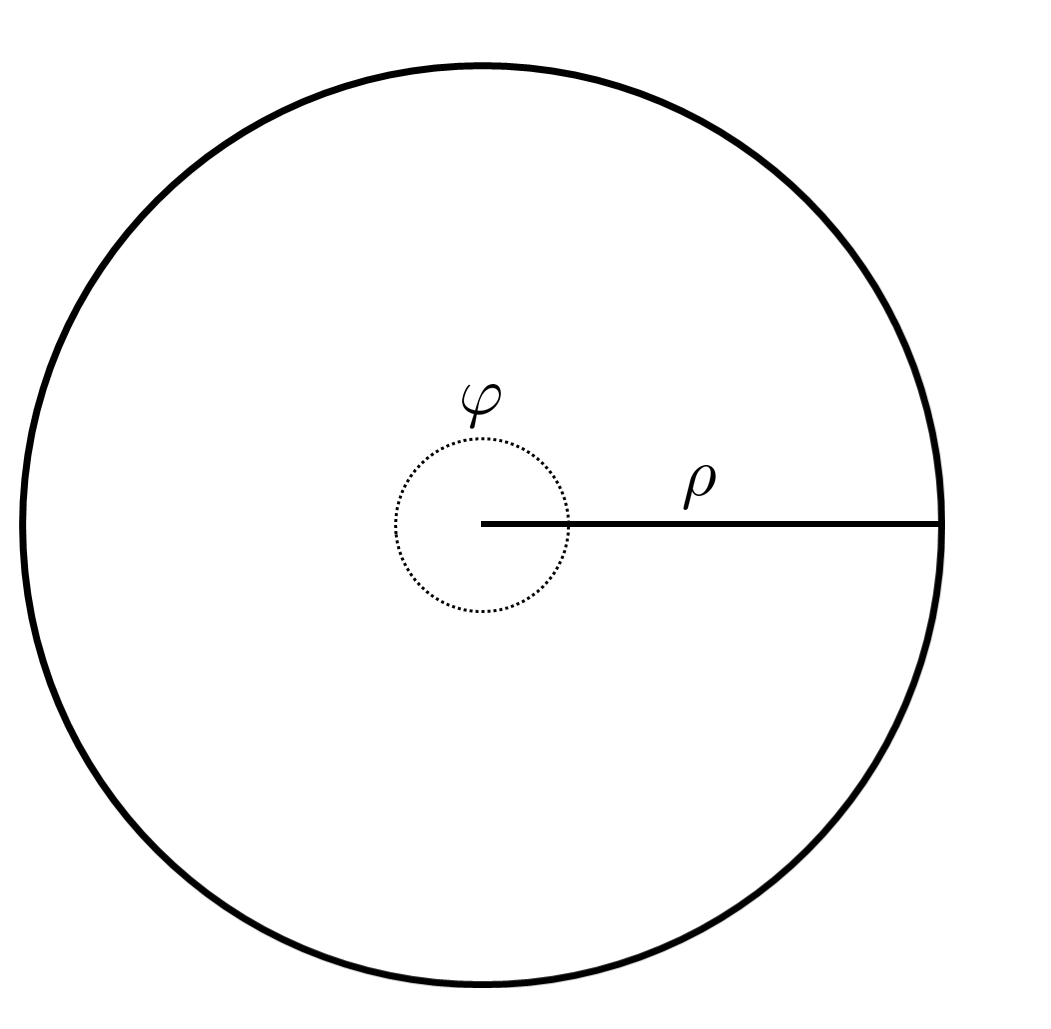} 
        \caption{} \label{Fig hyperbolic disk}
    \end{subfigure}
    \hspace{1cm}
    \begin{subfigure}[t]{0.45\textwidth}
        \centering
        \includegraphics[width=\linewidth]{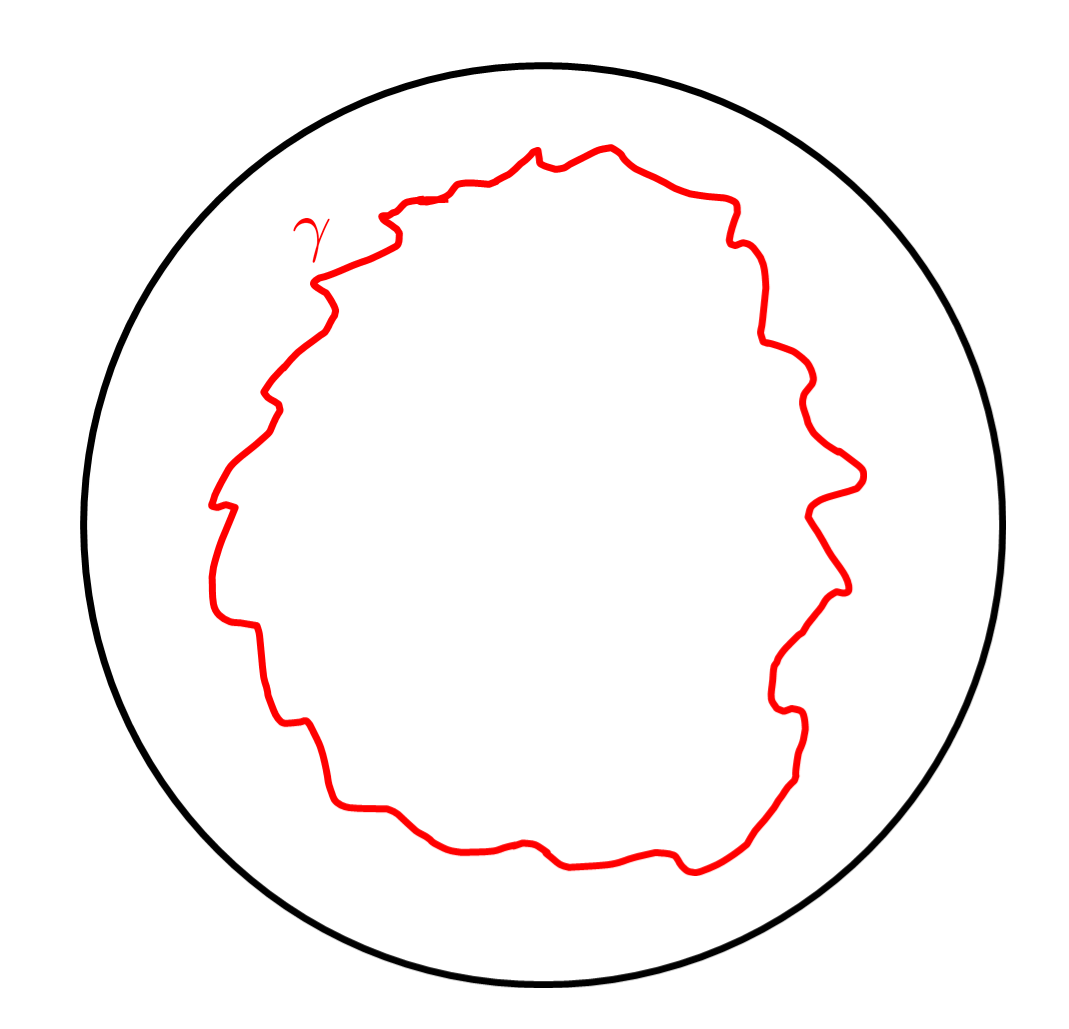} 
        \caption{} \label{Fig boundary wiggles}
    \end{subfigure}
    \caption{Figure \ref{Fig hyperbolic disk} shows the hyperbolic plane in Poincaré coordinates. These are polar-like coordinates, with a radius $\rho$, and an angle $\varphi$, which make the disk topology apparent. Figure \ref{Fig boundary wiggles} illustrates the calculation of the disk partition function of JT gravity. The path integral has to be performed over all boundary paths $\gamma$ with a fixed length. This figure shows one path that contributes to the integral.}
\end{figure} 
Starting from AdS$_{2+1}$, the hyperbolic plane $\mathds{H}^2$ may be regarded as a Cauchy slice of the three-dimensional spacetime. In global coordinates $\{\rho,t,\varphi\}\in[0,1[\times\mathds{R}\times[0,2\pi[$ this leads to the Poincaré disk, with induced line element
\begin{equation}
    ds^2=(2L_{\text{AdS}})^2\frac{d\rho^2+\rho^2d\varphi^2}{(1-\rho^2)^2}\,,
    \label{eq:Disk coordinates}
\end{equation}
where $L_{\text{AdS}}$ is the AdS-radius, which is related to the Ricci curvature by $R=-2/L_{\text{AdS}}^2$. The boundary of the hyperbolic plane is located at $\rho\to 1$. The hyperbolic plane in Poincaré disk coordinates is shown in Fig.~\ref{Fig hyperbolic disk}. Alternatively, the hyperbolic plane $\mathds{H}^2$ can also be seen as the Euclidean version of two-dimensional AdS, namely EAdS$_2$. JT gravity is a two-dimensional theory of gravity, with a metric $g_{\mu\nu}$ and a dilaton $\phi$, with Euclidean action \cite{JACKIW1985343, TEITELBOIM198341, MaldacenaYang2016}
\begin{equation}
S_{\text{JT}}=-\frac{\phi_0}{4G}\chi(\mathcal{M})-\frac1{16\pi G}\kd{\int_\mathcal{M} d^2x\sqrt{g}\phi(R+2/L_{\rm AdS}^2) 
    +2\int_{\partial \mathcal{M}} du\sqrt{h}\phi (K-1/L_{\rm AdS})}\,,
    \label{eq:conti JT}
\end{equation}
where $\phi_0$ is the background dilaton, $h$ is the determinant of the boundary metric, and $K$ is the extrinsic curvature. Furthermore $\chi(\mathcal{M})$ denotes the Euler characteristic of the underlying manifold, which is given by $\chi(\mathcal{M})=2-2g-n$, where $g$ is the genus of $\mathcal{M}$, and $n$ is the number of boundaries. The generating functional of JT gravity is given by 
\begin{equation}
    Z=\int\frac{\mathcal{D}g_{\mu\nu}\mathcal{D}\phi}{\text{Diffeos}}e^{-S_{\text{JT}}}\,.
    \label{eq:JT PI}
\end{equation}
Integrating out the dilaton fixes the curvature of the manifold to $R=-2/L^2_{\text{AdS}}$, i.e.~the geometry has to be a patch of the hyperbolic plane.  This significantly simplifies the path integral, since only metrics with constant negative curvature contribute, which leads to
\begin{equation}
    Z=e^{\frac{\phi_0}{4G}(2-2g-n)}\int d\Omega\int\mathcal{D}\gamma\exp\left(\frac{1}{8\pi G}\int_{\partial \mathcal{M}} du\sqrt{h}\phi (K-1/L_{\rm AdS})\right)\,,
\end{equation}
where $d\Omega$ denotes an integral over the moduli space of $\mathcal{M}$, while $\mathcal{D}\gamma$ is a path integral over boundary paths \cite{Saad2019}. For the remainder of this paper, we set $L_{\rm AdS}=1$. The most commonly studied quantity in the context of holography is the disk partition function. This may be regarded as the leading order in the topological expansion of the JT path integral, which corresponds to the semiclassical limit $G\to0$ \cite{MaldacenaYang2016,Saad2019,Stanford2020}. The moduli space of the disk is empty, which further simplifies the calculation to a path integral over all boundary paths, as is illustrated in Fig.~\ref{Fig boundary wiggles}. We impose boundary conditions
\begin{equation}
    \abs{\gamma}=\mathcal{L}\,,\qquad\phi\vert_\gamma=\phi_b\,,
    \label{eq:Boundary condition}
\end{equation}
where $\mathcal{L}$ is the boundary length, and $\phi_b$ is the boundary value of the dilaton. We evaluate the disk partition function by using the Gauß-Bonnet theorem
\begin{equation}
    \int_{\partial \mathcal{M}} du\sqrt{h}K=2\pi\chi(\mathcal{M})-\frac{1}{2}\int_\mathcal{M}d^2x\sqrt{g}R=2\pi+\text{Vol}(\mathcal{M})\,,
    \label{eq:Conti Gauß-Bonnet}
\end{equation}
where in the second equation we used that for the disk $(g,n)=(0,1)$ and $R=-2$. This yields
\begin{equation}
    Z_{\text{disk}}(\mathcal{L})\sim\int\mathcal{D}\gamma\,e^{\frac{\phi_b}{8\pi G}(A_\gamma-\abs{\gamma})}\delta\left(\abs{\gamma}-\mathcal{L}\right)\,,
    \label{eq:disk partition function}
\end{equation}
where $\sim$ implies that we dropped a proportionality constant, and $A_\gamma$ is the area enclosed by $\gamma$ \cite{Yang2018,Stanford2020}. The path integral \eqref{eq:disk partition function} can explicitly be solved in the Schwarzian limit, which corresponds to taking $G\to0$, as well as $\mathcal{L}\to\infty$. This may be viewed as taking a saddle point approximation of \eqref{eq:disk partition function}, and the disk partition function is given by
\begin{equation}
    Z_{\text{disk}}(\mathcal{L})=e^{S_0}\frac{C^{3/2}}{(2\pi)^{1/2}\mathcal{L}^{3/2}}e^{\frac{2\pi^2C}{\mathcal{L}}}\,, 
    \label{eq:JT partition function}
\end{equation}
where we set $C=\phi_b/8\pi G$. We also reintroduced the proportionality constant, which is given by $S_0=\phi_0/4G$.  JT gravity has also been studied at finite boundary lengths away from the Schwarzian regime \cite{Stanford2020,Iliesiu:2020zld}. We may think of $Z_{\text{disk}}(\mathcal{L})$ as a thermal partition function, which is related to the density of states by a Laplace transformation
\begin{equation}
    Z_{\text{disk}}(\mathcal{L})=\int_0^\infty dE\,e^{-\mathcal{L}E}\rho_0(E)=e^{S_0}\frac{C}{2\pi^2}\int_0^\infty dE\,e^{-\mathcal{L}E}\sinh(2\pi\sqrt{2C E})\,.
    \label{eq:Laplace trafo partition function}
\end{equation}
The duality between JT gravity and a matrix model arises by identifying the density of states, with the leading eigenvalue density of the dual theory \cite{Saad2019}.

\subsection{Hyperbolic lattices and inflation rules}
\label{subsec:inflationTiling}
\begin{figure}[t]
    \centering
    \begin{subfigure}[t]{0.45\textwidth}
        \centering
\includegraphics[width=\linewidth]{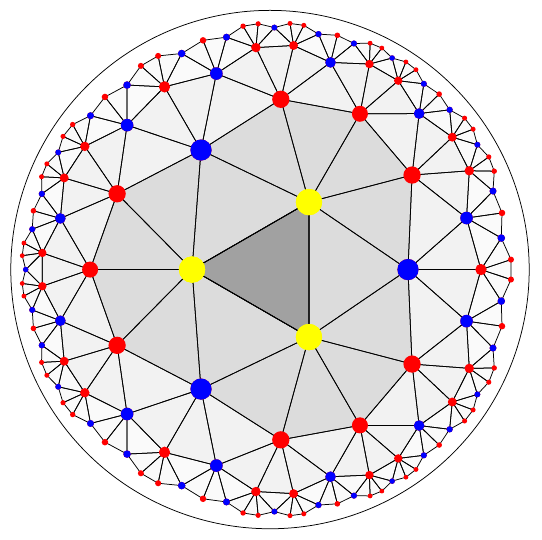} 
        \caption{} \label{Fig 3,7 tiling}
    \end{subfigure}
    \hspace{1cm}
    \begin{subfigure}[t]{0.45\textwidth}
        \centering
        \includegraphics[width=\linewidth]{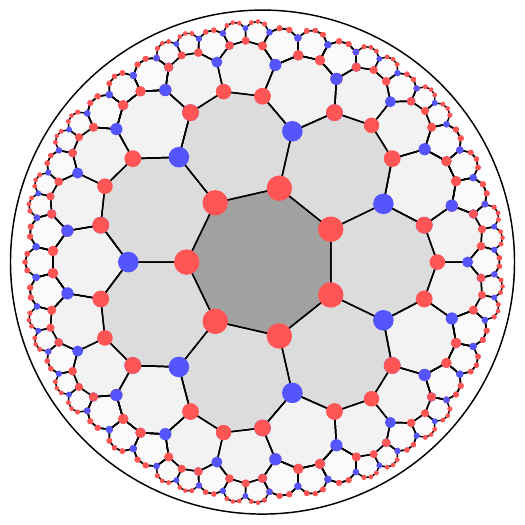} 
        \caption{} \label{Fig 7,3 tiling}
    \end{subfigure}
    \caption{Different examples of hyperbolic tilings. The tilings are constructed layer by layer via an inflation procedure. The different shades of gray indicate successive layers, while the red, blue, and yellow vertices have associated letters $a,b,c$ as defined by the inflation procedure. Figure \ref{Fig 3,7 tiling} shows a \pq{3}{7} tiling, while Figure \ref{Fig 7,3 tiling} 
    shows a \pq{7}{3} tiling, where for each tiling four layers have been constructed. Figure \ref{Fig 3,7 tiling} is taken from \cite{Basteiro2022}.}
\end{figure} 
We introduce the canonical discretization of the Poincaré disk via hyperbolic tilings \cite{Magnus1974, coxeter_1997}. These tilings are defined as gapless fillings of the hyperbolic space via regular hyperbolic polygons. In two dimensions, general tilings are determined by their Schläfli symbol \pq{p}{q}, which states that the tiling is obtained from regular $p$-gons, with $q$ polygons meeting at each vertex. In order to tessellate the hyperbolic plane, the parameters $p$ and $q$ have to satisfy $(p-2)(q-2)>4$, which admits an infinite number of solutions. For the purpose of this paper, we consider \pq{3}{q} tilings, as well as their dual tiling \pq{q}{3}, where we have $q\geq7$. The dual tiling is obtained by replacing each vertex in the original tiling with a polygon, and vice versa. Some examples of hyperbolic \pq{p}{q} tilings are shown in Fig.~\ref{Fig 3,7 tiling}-\ref{Fig 7,3 tiling}. We explain a construction mechanism of hyperbolic tilings via so-called inflation rules. A \pq{3}{q} tiling may be constructed starting from a central triangle, and adding concentric layers of tiles towards the boundary \cite{FlickerBoyle}. This construction mechanism can be used to define two types of vertices $a$ and $b$ \cite{Jahn:2019mbb,Basteiro2022}. A vertex with the associated letter $a$ is connected to one vertex in the previous layer, while a vertex with the associated letter $b$ is connected to two vertices in the previous layer. We have to add a third letter $c$, which is associated with the vertices of the central triangle since they have zero neighbors in the "previous" layer. Therefore, the central triangle corresponds to a seed word $ccc$. The sequence of letters in the $(l+1)$-th layer can be determined from the letter sequence in the $l$-th layer via a so-called inflation rule \cite{Jahn:2020ukq,Jahn:2019mbb,Basteiro2022}

\begin{equation}
\begin{split}
\sigma_{\{3,q\}}=\begin{cases}
a\mapsto a^{q-5}b\,,\\
b\mapsto a^{q-6}b\,,\\
c \mapsto a^{q-4}b\,.
\end{cases}
\end{split}
\label{InflationRulesForP=3}
\end{equation}
We note that the letter $c$ drops out after the first inflation step since it is only associated with the vertices of the central triangle. For the dual tiling, the same construction algorithm holds \cite{Basteiro2022}. Here the letter $a$ indicates that the vertex is not connected to the previous layer. A vertex with the associated letter $b$ is connected to one vertex in the previous layer. Since the original \pq{3}{q} tiling is centered around a triangle, the dual tiling is centered around a vertex, and again we associate a separate letter $c$ to it. The inflation rule for the \pq{q}{3} tiling is given by \cite{Jahn:2020ukq,Jahn:2019mbb,Basteiro2022,Basteiro2022a}
\begin{equation}
\begin{split}
\sigma_{\{q,3\}}=\begin{cases}
a\mapsto a^{q-4}b \,,\\
b\mapsto a^{-1}\,,\\
c \mapsto (a^{q-3}b)^3\,,
\end{cases}
\end{split}
\label{InflationRulesDualTiling}
\end{equation}
where the letter $c$ again drops out after the first inflation step. The construction of a hyperbolic tiling via inflation rules is illustrated in Fig.~\ref{Fig 3,7 tiling}-\ref{Fig 7,3 tiling}. Since the sequence of letters on the asymptotic boundary of the tiling consists of two letters it is called binary. The properties of this sequence are encoded in the substitution matrix, which reads

\begin{equation}
M_{\{3,q\}}=\left(
\begin{matrix}
q-5 & q-6 \\
1 & 1 \\
\end{matrix}
\right)\,,\qquad
M_{\{q,3\}}=\left(
\begin{matrix}
q-4 & -1 \\
1 & 0 \\
\end{matrix}
\right)\,.
\label{eq:InflationMatrix}
\end{equation}
The entries $(M_{\{p,q\}})_{ij}$ with $i,j\in\{a,b\}$ are the number of times the letter $i$ appears in the substitution word for the letter $j$ as defined by  \eqref{InflationRulesForP=3} and \eqref{InflationRulesDualTiling}. Since this matrix is real and positive by construction, there exists a unique largest eigenvalue, which for both matrices is given by

\begin{equation}
\lambda_+=
\frac{1}{2}\left(q-4+\sqrt{q^2-8q+12}\right)\,.
\label{eq:LargestEV}
\end{equation}
The asymptotic scaling of the number of boundary vertices $V_\pa$ after a large number of $l\gg1$ inflation steps is encoded in this eigenvalue as $V_\pa\sim\lambda_+^l$. Thus, it is important to note that $\lambda_+>1$. The components of the right eigenvector, $\mathbf{v}_+=(p_a,p_b)^T$ with $p_a+p_b=1$, count how frequently letters $a,b$ i appear in the asymptotic sequence. For the left eigenvector, $\mathbf{u}_+=(l_a,l_b)^T$ with $\mathbf{u}_+\cdot\mathbf{v}_+=1$, the components measure  the typical length of the letter sequences. 

\subsection{JT gravity on a hyperbolic lattice}

We proceed to define JT gravity on a hyperbolic lattice. A hyperbolic tiling provides a smooth discretization of the hyperbolic space. This is different from the discretization scheme used in lattice approaches to quantum gravity. There the general idea is to replace the smooth spacetime manifold $\mathcal{M}$ by a simplicial complex $\Delta$ \cite{Regge1961}. This complex is viewed as a discrete approximation of $\mathcal{M}$. In two dimensions, a simplicial complex is a lattice built from flat triangles. This is different from a hyperbolic tiling, where the triangles themselves are hyperbolic, and  thus possess curvature. For a simplicial complex, curvature is induced at the vertices through the deficit angle
\begin{equation}
    \varepsilon_v=2\pi-\sum_{\sigma\subset v}\theta(\sigma,v)\,,
    \label{eq:deficit angle}
\end{equation}
where $\sigma$ denotes the triangles meeting at the vertex $v$ and $\theta(\sigma,v)$ is the internal angle of a triangle at that vertex. As an example, consider equilateral triangles, which have internal angles $\theta=\pi/3$. If six triangles meet at a vertex, then $\varepsilon_v=0$. Adding or removing a triangle induces a positive/negative deficit angle, i.e.~curvature at each vertex. Thus, the Ricci scalar and the extrinsic curvature of a simplicial complex are given by
\begin{equation}
    R(x)=\sum_{v\in\Delta^\circ} 2\varepsilon_v\,\delta^{(2)}(x-v),\quad K(x)=\sum_{v\in\partial \Delta} \psi_v\,\delta^{(1)}(x-v)\,,
    \label{eq:Average curvature}
\end{equation}
where $\psi_v$ is the deficit angle at the boundary \cite{Hartle:1981}. Also, $\Delta^\circ$ denotes the set of internal vertices, while $\partial \Delta$ is the set of boundary vertices. The volume of the simplicial complex is given by
\begin{equation}
    \text{Vol}(\Delta)=\sum_{v\in  \Delta^\circ}a_v\,,
    \label{eq:Volume complex}
\end{equation}
where $a_v$ is $1/3$ of the total area of the triangles sharing a vertex $v$. This term may be used to introduce a cosmological constant to the discrete gravity action. We consider JT gravity on a $\ke{3,q}$ lattice, i.e.~a lattice $\Delta$ built up from equilateral triangles, where $q$ triangles meet at each vertex. For a hyperbolic lattice, we have $q\geq7$. We obtain the Regge action of JT gravity by inserting \eqref{eq:Average curvature} into \eqref{eq:conti JT}, which leads to 
\begin{equation}
    S_R=-\frac{\phi_0}{4G}\chi(\Delta)-\frac1{8\pi G}\kd{ \sum_{v\in  \Delta^\circ}\phi_v(\varepsilon_v+a_v)
    +\sum_{v\in\partial \Delta}\phi_v(\psi_v-\psi_c)}\,,
    \label{eq:Discrete JT action}
\end{equation}
where $\psi_c$ is the average defect angle on the conformal boundary. This angle acts as the counterterm, which is derived in appendix \ref{apx:counterterm}, and given by \eqref{eq:counterterm}. The Euler characteristic of a simplicial complex is given by $\chi(\Delta)=V-\mathcal{E}+F$, where $V,\mathcal{E},F$ are the number of vertices, edges, and faces in $\Delta$. The discrete gravity path integral is given by 
\begin{equation}
Z=\int\mathcal{D}l_{ij}^2\mathcal{D}\phi\,e^{-S_R}\,,\\
    \label{eq:path integral}
\end{equation}
 where we integrate over the dilaton, as well as the edge fluctuations of the triangles as defined in \cite{Hamber2007}.  Integrating out the dilaton leads to
\begin{align}
0=\varepsilon_v+a_v\,.
\end{align}
This fixes the simplicial geometry as discrete hyperbolic space with average curvature $\overline{R}=-2$, since
\begin{align}
    \frac12 \overline{R} \int d^2x\sqrt{g}=\frac12\int d^2x\sqrt{g} R=\sum_v \varepsilon_v=-\sum_v a_v=-\int d^2x\sqrt{g}\,.
\end{align}
\noindent Since we consider a $\ke{3,q}$ triangulation, each triangle is equilateral with angle $\theta=\pi/3$, lower edge length $s$, and area $a_\Delta$. We obtain
\begin{align}
    \label{eq:sizes of triangle}
    \left(2-\frac q3\right)\pi=\varepsilon=-a=-\frac{qa_\Delta}{3}=-\frac{qs^{2}}{4\sqrt3}\,,\quad q\geq 7\,.
\end{align}
This fixes the edge length of each triangle, and the path integral \eqref{eq:path integral} reduces to an integral over boundary fluctuations, as in the continuous case. In the classical limit $G\to0$ contributions from higher topologies are suppressed and we consider discrete JT gravity on the disk. We fix a constant boundary dilaton $\phi\vert_{\partial\Delta}=\phi_b$, as well as the boundary length $\mathcal{L}$, as in the continuous case \eqref{eq:Boundary condition}. We therefore have to sum over all possible discrete boundary paths $\Gamma$ with fixed length that consist of triangle edges.  The disk partition function is thus obtained from \eqref{eq:path integral} as
\begin{equation}   Z(\mathcal{L})=\text{e}^{\frac{\phi_0}{4G}\chi(\Delta)}\sum\limits_{\abs{\Gamma}=\mathcal{L}}\exp\left(\frac{\phi_b}{8\pi G}\sum\limits_{v\in\partial \Delta}(\psi_v-\psi_c)\right)\,.
    \label{eq:discrete disk partition function}
\end{equation}
The sum can be evaluated by using the discrete analogue of the Gauß-Bonnet theorem,
\begin{align}
\chi(\Delta) =\frac1{2\pi}\sum_{v\in \Delta^\circ}\varepsilon_v +\frac1{2\pi}\sum_{v\in\partial \Delta}\psi_v
    =\frac1{2\pi}\varepsilon V_\tint+\frac1{2\pi}\sum_{v\in\partial \Delta}\psi_v\,,
    \label{eq:discrete Gauß Bonnet}
\end{align}
where we define the number of interior vertices as $V_\tint=\sum_{v\in \Delta^\circ}1$. The number of boundary vertices is thus given by $V_\partial=\sum_{v\in\pa \Delta}1=V-V_\tint$. We find
\begin{equation}
    \begin{split}
        Z(\mathcal{L})&=e^{\frac{\phi_0+\phi_b}{4G}}\sum\limits_{\abs{\Gamma}=\mathcal{L}}\exp\left[-\frac{\phi_b}{8\pi G}\left(\sum\limits_{v\in \Delta^\circ}\varepsilon_v+\psi_cV_\partial\right)\right]\\
        &=e^{\frac{\phi_0+\phi_b}{4G}}\sum\limits_{\abs{\Gamma}=\mathcal{L}}\exp\left[\frac{\phi_b}{8\pi G}\left(\sum\limits_{v\in \Delta^\circ}a_v-\psi_cV_\partial\right)\right]=e^{\frac{\phi_0+\phi_b}{4G}}\sum\limits_{\abs{\Gamma}=\mathcal{L}}\text{e}^{\frac{\phi_b}{8\pi G}(A_\Gamma-\psi_cV_\partial)}\,,
    \end{split}
    \label{eq:Calculation Disk}
\end{equation}
\noindent where $A_\Gamma$ is the area enclosed by $\Gamma$. This equation is the discrete analog of \eqref{eq:disk partition function}. Instead of integrating over all possible boundary paths, we have to sum over discrete paths, which consist of the edges of the triangles. In the next section, we show that this calculation is equivalent to evaluating the partition function of an Ising model. This allows us to simulate discrete JT gravity, and to evaluate the path integral.

\section{Discrete JT gravity as an Ising model}
\label{sec:Ising model}

To further study discrete JT gravity, we now show that there exists a one-to-one correspondence between this gravity theory and an Ising model on the dual lattice. We use a Monte Carlo approach to simulate the Ising model and obtain its free energy, which is related to the resolvent of JT gravity. 

\subsection{Mapping the bulk theory to an Ising model}
\label{subsec:Mapping}

\begin{figure}
    \centering
    \includegraphics[width=1.0\textwidth]{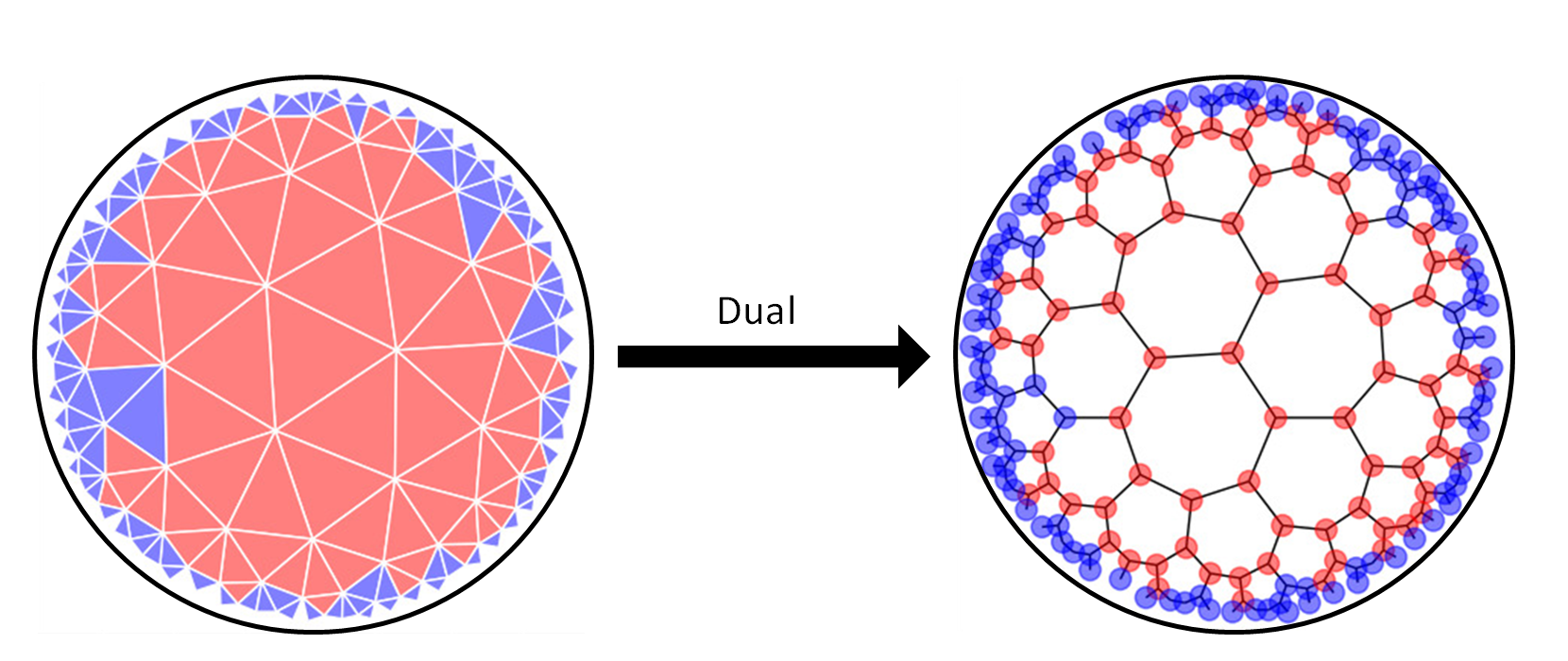} 
        \caption{Illustration of the map from the discrete JT gravity theory to an Ising model. On the left figure, we show one geometry contributing to \eqref{eq:Calculation Disk}. The area shown in red is bounded by a closed path of a given length on the hyperbolic lattice that contributes to the discretized path integral.  The region outside of the path is shown in blue.
        Each triangle is mapped to a vertex of the dual \pq{7}{3} tiling on the right. A red triangle is interpreted as a spin-up, while blue triangles are interpreted as a spin-down. Thus the discretized path integral \eqref{eq:Calculation Disk} may be viewed as a sum over spin configurations on the dual lattice, with a fixed length of the domain wall.} \label{Fig:Map to Ising}
   
\end{figure} 
We show that evaluating the sum \eqref{eq:Calculation Disk} is equivalent to the partition function of an Ising model. For the $\ke{3,q}$ triangulation considered here, we have
$V_\pa=2\mathcal{E}-3F$. Starting from \eqref{eq:Calculation Disk}, we obtain
\begin{equation}
    \begin{split}
         S_R
    =&~-\myroundedbrackets{\frac{\phi_0+\phi_b}{4G}-\frac{\phi_b\,\varepsilon}{8\pi G}}\chi(\Delta)+\frac{\phi_b\,\varepsilon}{16\pi G}F+\frac{\phi_b(2\psi_c-\varepsilon)}{16\pi G}V_\pa\\
    =&~-\beta\myroundedbrackets{\gamma_\phi+\frac{q}{6}}\chi(\Delta)
    -\beta h\sum_{r\in \Delta^\times}\frac{s_r+1}2
    -\beta J\sum_{\avg{rr'}\in  \Delta^\times}\frac{s_rs_{r'}-1}2\,,
    \end{split}
    \label{eq:Mapping to Ising model}
\end{equation}
\noindent with spin $s_r=\pm1$,  and parameters
\begin{equation}
\gamma_\phi=\frac{\phi_0}{\phi_b},\quad 
    \beta=\frac{\phi_b}{4 G},\quad
    h=-\frac{\varepsilon}{4\pi}=\frac{q-6}{12},\quad
    J=\frac{2\psi_c-\varepsilon}{4\pi}=\frac{1}{12} \sqrt{(q-6) (q-2)}\,,
    \label{eq:Magnetic field and coupling constant}
\end{equation}
\noindent where $\phi_b,\,\varepsilon,$ and $\psi_c$ are given by \eqref{eq:Boundary condition}, \eqref{eq:sizes of triangle}, and \eqref{eq:counterterm}. These parameters may be interpreted as an inverse temperature $\beta=T^{-1}$, a magnetic field $h$, and a coupling constant $J$. Also, $\phi_0$ is the background dilaton, $\chi(\Delta)$ is the Euler characteristic of the lattice $\Delta$, while $\Delta^\times$ denotes the dual lattice, i.e.~$\ke{q,3}$, which appears in the sum over spin variables. The logic behind this mapping is as follows. In \eqref{eq:Calculation Disk} we sum over all closed paths of the entire $\ke{3,q}$ lattice, with a fixed length, and weigh them by the area of the region enclosed by the path. We assign a spin up to each triangle enclosed by the path while assigning a spin down to all other triangles. Thus the number of spin-ups corresponds to the number of faces inside the path, i.e.~$F=\frac{1}{2}\sum_r(s_r+1)$. The number of boundary vertices corresponds to the number of neighboring spins with opposite signs $V_\pa=-\frac12\sum_{\avg{rr'}}(s_rs_{r'}-1)$, where $\avg{rr'}$ refers to nearest neighboring sites, connected by a link. Instead of summing over paths, \eqref{eq:Calculation Disk} becomes a sum over spin configurations on the dual lattice, with fixed boundary length of the domain wall $\mathcal{L}$. This length is proportional to the number of boundary vertices $V_\pa$, and reads $\mathcal{L}=\beta V_\pa$ since we work in dimensions of the dilaton. Due to the fact that the paths have fixed finite lengths we have to impose the boundary condition $s_\infty=-1$. This mapping is illustrated in Fig.~\ref{Fig:Map to Ising}. To obtain the disk partition function of discrete JT gravity, we have to fix the Euler characteristic to the value of the disk
\begin{equation}
    \begin{split}
        1=&~\chi(\Delta)=V-\mathcal{E}+F=\frac12(V_\pa-F+2V_\tint)\, ,\\
    \end{split}
    \label{eq:Disk topology}
\end{equation}
where $V,\mathcal{E},F$ is the number of vertices, edges, and faces in the lattice while $V_\partial$ is the number of boundary vertices, and $V_\tint$ is the number of bulk vertices. The last equation is a consequence of the fact that we consider a \pq{3}{q} lattice, which satisfies $V_\pa=2\mathcal{E}-3F$. The number of bulk vertices of the original \pq{3}{q} lattice is given by the number of faces on the dual \pq{q}{3} lattice. This corresponds to the number of closed loops of spin ups, i.e. $2^{1-q}\sum_{\avg{r_1r_2\cdots r_q}}(s_{r_1}+1)(s_{r_2}+1)\cdots(s_{r_q}+1)$. In terms of the spin variables the constraint \eqref{eq:Disk topology} reads
\begin{equation}
     1=\frac12\kc{-\sum_r \frac{s_r+1}2-\sum_{\avg{rr'}}\frac{s_rs_{r'}-1}2+2\sum_{\avg{r_1r_2\cdots r_q}}\frac{s_{r_1}+1}2\frac{s_{r_2}+1}2\cdots\frac{s_{r_q}+1}2}\,,
     \label{eq:Constraint disk topology}
\end{equation}
where $\avg{r_1r_2\cdots r_q}$ refers to the sites forming a $q$-gon. This constraint ensures that the domain of the spins pointing up has no holes. The partition function of JT gravity \eqref{eq:Calculation Disk} becomes
\begin{align}
    Z(\mathcal{L})=\sum_{\ke{s}}e^{-S_R}\,\delta(V_\pa \beta-\mathcal{L})\,,
    \label{eq:result discrete disk partition}
\end{align}
where we sum over all spin configurations that obey the topological constraint \eqref{eq:Constraint disk topology}. In order to evaluate the discrete disk partition function numerically, we integrate out the contribution from the delta constraint in \eqref{eq:result discrete disk partition}. This leads to the resolvent function $R(E)$, which is related to the partition function by
\begin{align}
    R(E)
    =-\int_0^\infty d\mathcal{L}\,e^{\mathcal{L} E}Z(\mathcal{L}) 
    =-\sum_{\ke{s}} e^{-\beta H}\,,
    \label{eq:Resolvent and partition function}
\end{align}
with
\begin{equation}
    \label{eq:Hamiltonian Ising model}
    \begin{split}
        H=&~S_R/\beta-EV_\pa
    =-\myroundedbrackets{\gamma_\phi+\frac{q}{6}}
    -h\sum_{r\in \Delta^\times}\frac{s_r+1}2
    -(J-E)\sum_{\avg{rr'}\in\Delta^\times}\frac{s_rs_{r'}-1}2\,.
    \end{split}
\end{equation}
This shows that the evaluation of the disk resolvent of JT gravity on a hyperbolic lattice is equivalent to calculating the partition function of an Ising model on the dual lattice. Thus $H$ given by \eqref{eq:Hamiltonian Ising model} may be interpreted as the Hamiltonian of the Ising model, with parameters given by \eqref{eq:Magnetic field and coupling constant}. We note that almost all parameters are fixed by $q$, i.e.~the underlying structure of the lattice. The only free parameters are $\gamma_\phi$, which corresponds to a shift of the ground state energy, as well as $E$, which appears in the coupling constant $J_{\text{eff}}=J-E$. Since $E$ is allowed to take arbitrary values the coupling can also take arbitrary positive, and negative values. Due to its origin from the disk partition function of discrete JT gravity, the Ising model is subject to a topological constraint \eqref{eq:Constraint disk topology}, which restricts the spins pointing up to form a single simply connected domain. This is a huge restriction on the space of allowed spin configurations and is very different from the usual Ising model. In the next section, we study the free energy of the Ising model numerically, to explicitly find the partition function of the model. We note that the effective description of JT gravity in terms of an Ising model requires an integral over all positive values for the length of the domain boundary $\mathcal{L}$. Due to  lattice finite size effects, it is not possible to approach the $\mathcal L\to\infty$ limit numerically. The relation between discrete JT gravity and the Ising model is only valid for proper values of $E$, or equivalently of the coupling constant $J_{\text{eff}}=J-E$, where the domain wall is away from the boundary of the hyperbolic lattice. 

\subsection{Simulation of the free energy}
\begin{figure}[!ht]
    \centering
    \includegraphics[width=\linewidth]{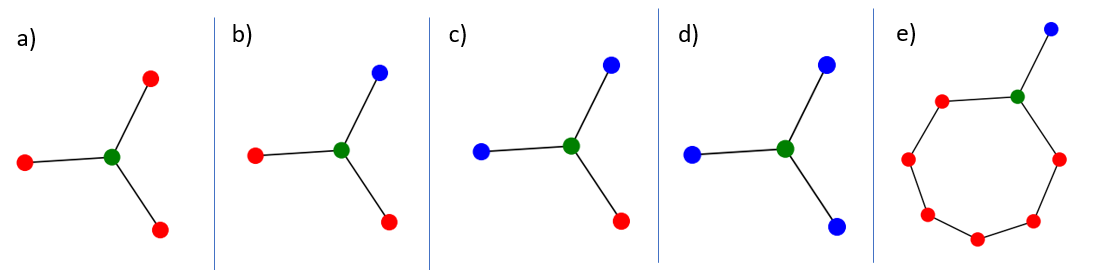}
    \caption{Possible spin configuration. The considered spin is green. Its neighbors either have the same (red) or a different (blue) orientation.}
    \label{fig:possible_spin_configs}
\end{figure}

The free energy of the Ising model is
\begin{equation}
    \mathcal{F}(E)=-\log(-R(E))\,,
    \label{eq:Ising Free energy}
\end{equation}
where the JT resolvent \eqref{eq:Resolvent and partition function} corresponds to the partition function of an Ising model with Hamiltonian \eqref{eq:Hamiltonian Ising model}. We note that here we define the free energy without the explicit prefactor $\beta^{-1}=T$ and therefore $\mathcal{F}$ is dimensionless. We use a Monte Carlo approach to simulate the free energy, where we introduce multistage sampling \cite{Smith1996}. The basic idea is to measure the difference in free energy between two canonical ensembles. Thus, consider two canonical ensembles at inverse temperature $\beta_0$ and $\beta_1$. The configurations in both systems follow a Boltzmann distribution, which may be written in terms of the energy $\epsilon$ of the underlying state as

\begin{equation}
    P_{i}(\epsilon)=n(\epsilon)\frac{e^{-\beta_i\epsilon}}{Z_{\text{can},i}}\,,
    \label{eq:Boltzmann energy distribution}
\end{equation}

\noindent where $i\in\{0,1\}$. The function $n(\epsilon)$ is the density of states of the thermodynamic system, and $Z_{\text{can},i}$ are the canonical partition functions of the two systems. Since $n(\epsilon)$ is unknown we cannot derive $Z_{\text{can}}$ explicitly, but instead consider

\begin{equation}
    \frac{Z_{\text{can},1}}{Z_{\text{can},2}}=\frac{P_{0}(\epsilon)}{P_{1}(\epsilon)}\exp(-(\beta_1-\beta_0)\epsilon)\,,
    \label{eq:ration partition function}
\end{equation}

\noindent which is used to estimate $Z_{\text{can},1}/Z_{\text{can},0}=\exp(\mathcal{F}_0-\mathcal{F}_1)$. If the probability distributions $P_{0}$ and $P_{1}$ overlap, we find

\begin{equation}
   \mathcal{F}_0-\mathcal{F}_1=\log\frac{\int_Sd\epsilon\,P_{0}(\epsilon)\exp(-(\beta_1-\beta_0)\epsilon)}{\int_Sd\epsilon\,P_{1}(\epsilon)}\,,
    \label{eq:difference free energy}
\end{equation}

\noindent where $S$ is the overlapping region of both probability densities \cite{Smith1996}. Choosing one of the two ensembles as a reference state with known free energy, we use \eqref{eq:difference free energy} to find the free energy of the other ensemble. In the simulation, this is implemented by decreasing the temperature of the system in each update step, after the system has been thermalized. By lowering the temperature we approach the ground state of the system. From the point of discrete JT gravity, lowering the temperature corresponds to taking the semiclassical limit $G\to 0$ \eqref{eq:Magnetic field and coupling constant}. Thus quantum gravity effects, such as contributions from other topologies, are suppressed. \\
When simulating the system we have to impose the constraint \eqref{eq:Constraint disk topology}, which restricts the topology of the Ising model states. As it would be highly unlikely to obtain a state with such conditions randomly, we enforce the topology to the Ising model by preventing a topology change first hand. As a \pq{q}{3} tiling is used,  the neighbors of a spin are in one of four possible states, relative to the spin considered. The possibilities are shown in Fig.~\ref{fig:possible_spin_configs}. The spin under consideration is depicted in green. Its neighbors either have the same (red) or a different (blue) orientation. However, for preserving the topology, not all of these spin configurations allow for the spin considered to be flipped. In a), all the neighbors have the same orientation as the considered spin. The considered spin is, therefore, not allowed to be flipped as it would directly violate the topological constraint. In b), two neighbors have the same orientation. This is the most difficult situation, which we will address later in more detail. In c), two neighbors are differently aligned. A flip of the spin is possible and can not change the topology. A similar situation holds for d), where all neighbors are differently aligned. This can only occur if the considered spin is the last spin of its type. In this case, we can, again, flip the spin. As mentioned above, b) is the most crucial case. In case the similarly aligned neighbors are only connected to the considered spin, a spin flip is not allowed. If, on the other hand, the spins are connected over the remaining spins, the spin can be flipped. Due to the topology, this connection has to be done within the closest loop connecting these spins e). This spin can be flipped if a loop connecting these two spins has a similar orientation. Therefore, for the update procedure, these loops need to be known. These can be detected by applying a bidirectional search for each pair of nodes, blocking the direct link.
\\
To construct the tiling, we use the hypertiling package developed at the JMU Würzburg \cite{Hypertiling,Schrauth:2023adh}. As we only need the graph, i.e. the connection of spins, we use GRGS. That is a static graph kernel provided by the package. 
Moreover, to improve performance, we apply the numba just-in-time compiler on the python-code to increase its performance. 

 We simulate the Ising model for $q=7$, i.e.~a lattice which consists of heptagons. For larger $q$, the number of spins in a chain increases. As the loops taking part in preserving the topology, the larger the loop, the lower the relative number of spins allowed to flip. This effectively reduces the update rate causing the simulation to requiere more computational time for finding the ground state. Therefore, we only consider a lattice with small $q$. The model is run with $\mathcal N=2.833$ changeable nodes, which corresponds to 14 layers. It is thermalized to its starting temperature $\beta=0$ with $2, 000, 000$ update steps, causing a random start configuration. Subsequently, it is cooled down from $\beta \in [0, 10]$ where the temperature is decreased in $300, 000$ steps. For each new $\beta$, the network performs $\mathcal N$ update steps. As mentioned at the end of the previous section, the parameter $E$ has to be fine-tuned to connect the Ising model to the original gravity theory. We sample $E \in [-30 \cdot J, 10 \cdot J]$ with $100$ steps, and use the numerical results to determine the appropriate regime. The code can be found at \cite{GITHUB}.

\subsection{Discussion of the numerical results}

\begin{figure}
   \centering
   \includegraphics[width=0.8\textwidth]{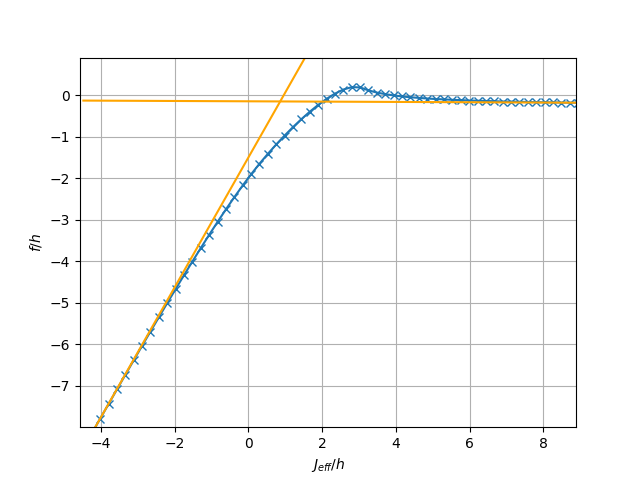}
        \caption{Free energy density $f=(\mathcal{F}/\beta+\gamma_\phi+q/6)/\mathcal N$ of the Ising model model with Hamiltonian \eqref{eq:Hamiltonian Ising model} as a function of the effective coupling $J_{\text{eff}}=J-E$  for a finite \pq{7}{3} lattice in units of the magnetic field $h$. The inverse temperature of the system is $\beta=T^{-1}=10$ and the number of nodes is $\mathcal{N}=2.833$. We identify two asymptotic regimes for the free energy. The magnetically dominated regime $J_{\text{eff}}\in]-\infty,J_{\text{min}}]$, in which the ground state of the system is given by a spin domain of maximal size. The energy of this spin configuration is given by a linear function in the coupling constant \eqref{eq:GS_maximal}. We plot a linear fit to the asymptotic points. The coupling-dominated regime $J_{\text{eff}}\in[J_{\text{max}},\infty[$, in which the size of the domain shrinks to zero, corresponds to the asymptotically constant behavior of the free energy \eqref{eq:GS_min}. We also fit a linear function to this asymptotic behavior. Additionally we find the expected intermediate regime, where the functional behavior of the free energy interpolates between the two asymptotic regimes. A necessary condition for the system to be in the ground state is that the free energy is below the asymptotic lines. The peak in the free energy is a numerical artifact, which we discuss in the text. In the interval between the numerical results leaving the linear fit on the left and before reaching the peak, the system corresponds to discrete JT gravity, realized by the Ising model with a finite-size domain.}
        \label{Fig:FreeEnergy}
\end{figure}

\begin{figure}[t]
    \centering
    \begin{tabular}{cc}
    \begin{subfigure}[t]{0.5\textwidth}
        \centering
        \includegraphics[width=\linewidth]{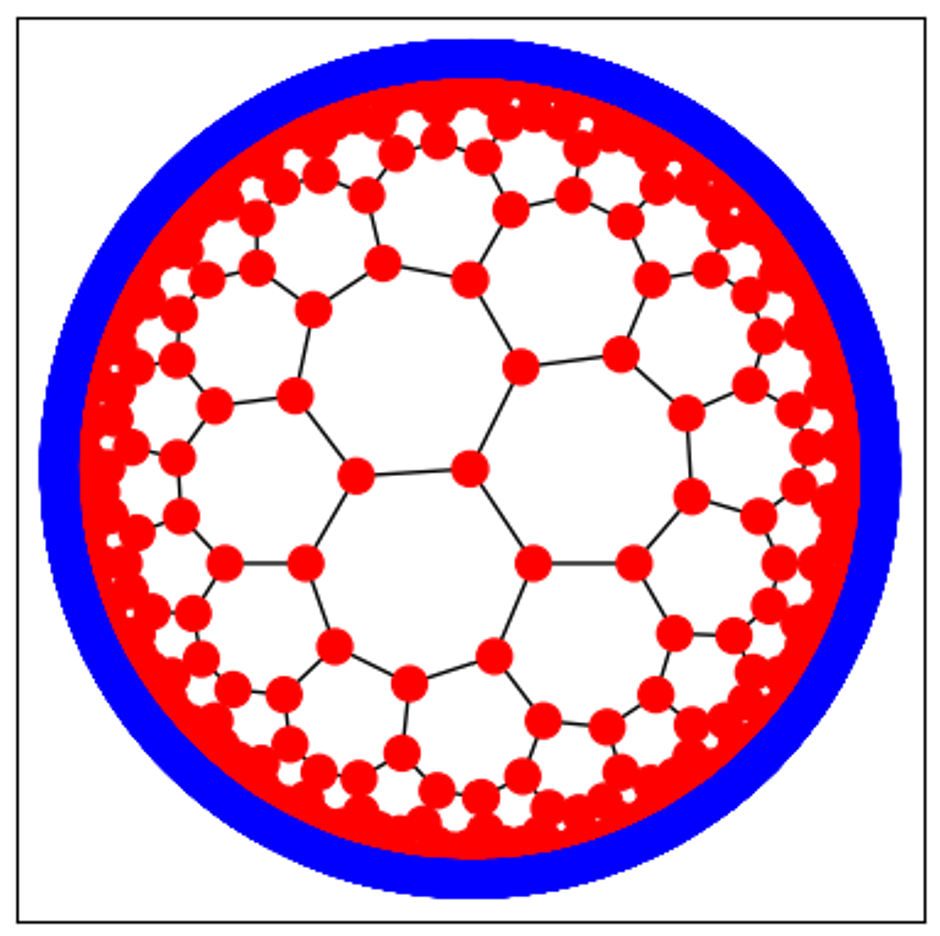} 
        \caption{} \label{Fig:Maximal domain}
    \end{subfigure}
    &
      \begin{subfigure}[t]{0.492\textwidth}
        \includegraphics[width=\linewidth]{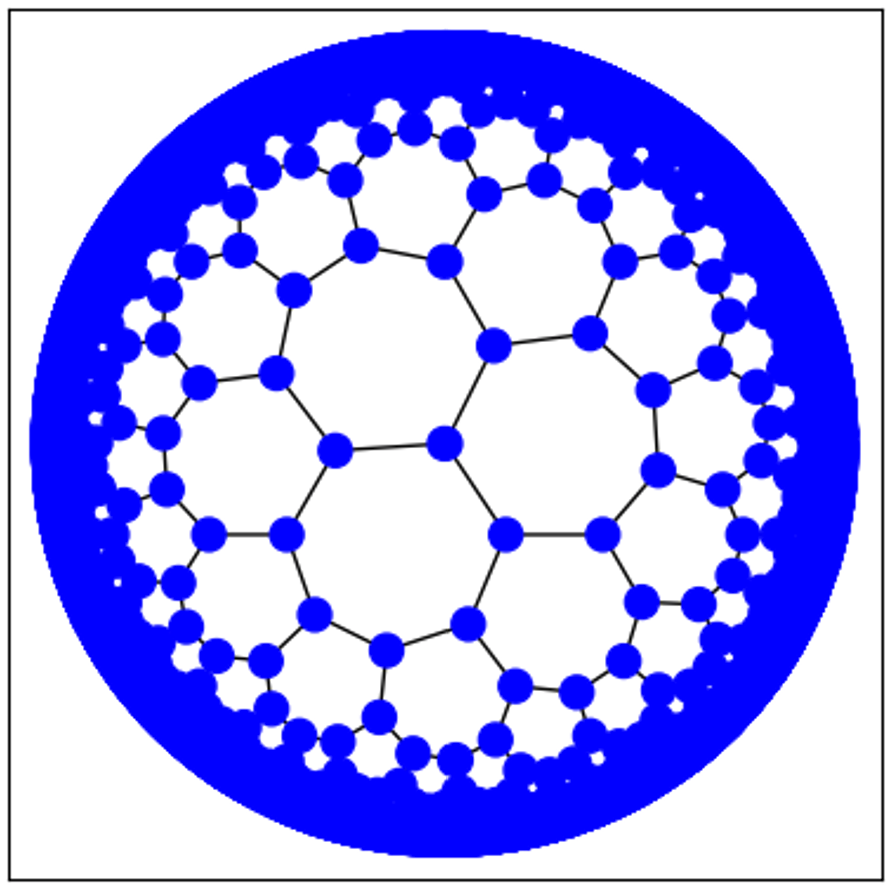} 
        \caption{} \label{Fig:Minimal domain}
    \end{subfigure}
    \end{tabular}
    \caption{Spin configuration of the Ising model for the different regimes of the parameter space. Figure \ref{Fig:Maximal domain} corresponds to the magnetically dominated regime, starting from a small positive coupling constant, and going to $J_{\text{eff}}\to-\infty$. Therefore \eqref{eq:Interpret Hamiltonian} is minimized, by maximizing the size of the domain, as can be seen in the figure. Figure \ref{Fig:Minimal domain} shows the coupling-dominated regime, which corresponds to a positive coupling constant that is large compared to the magnetic field. In this regime the Hamiltonian \eqref{eq:Interpret Hamiltonian} is minimized, by minimizing the size of the domain. This leads to all spins pointing down, which can be seen in the figure.}
\end{figure}

The free energy density obtained from the numerical simulation explained in the previous section is shown in Fig.~\ref{Fig:FreeEnergy}, as a function of the effective coupling $J_{\text{eff}}=J-E$ of the Hamiltonian \eqref{eq:Hamiltonian Ising model} in units of the magnetic field. In the plot, we focus on the intermediate regime and do not show the full range of the asymptotic behavior as it is linear. The free energy \eqref{eq:Ising Free energy} is related to the internal energy $U$ and the entropy $S$ by
\begin{equation}
    \mathcal{F}=\frac{1}{T} U-S\,.
    \label{eq:Free energy, internal energy}
\end{equation}
In the limit $T\to 0$ this implies $\mathcal{F}\approx T^{-1}\cdot U$. Thus the free energy can be approximated by the energy of the ground state, i.e.~the minimum of the Hamiltonian. We note that due to the topological constraint \eqref{eq:Constraint disk topology}, as well as the boundary condition $s_\infty=-1$ the ground state is not degenerate. In particular, we do not observe geometric frustration for negative coupling constants \cite{jalagekar2023geometric}. We rewrite the Hamiltonian \eqref{eq:Hamiltonian Ising model} as 
\begin{equation}
    H=-\myroundedbrackets{\gamma_\phi+\frac{q}{6}}-hF+J_{\text{eff}}V_\pa\,,
    \label{eq:Interpret Hamiltonian}
\end{equation}
\noindent where $F$ denotes the number of spins pointing up, while $V_\pa$ is the number of neighboring spins with opposite signs. The constant in front of the Hamiltonian determines the energy of the ground state and is obtained from the free parameter $\gamma_\phi$ of the gravity theory. We can set the constant to zero, without loss of generality. Due to the topological constraint \eqref{eq:Disk topology} from the gravity path integral, only spin configurations where the spins pointing up form a simply connected domain are allowed. From the point of view of standard Ising models, this is of course a very special case. For the simply connected domain, $F$ is proportional to the size of the domain, while $V_\pa$ is proportional to the length of its boundary  (see Fig.~\ref{Fig:Map to Ising}). When calculating the free energy, we minimize the Hamiltonian \eqref{eq:Interpret Hamiltonian}. We find three different regimes in the phase structure of the Ising model, depending on the ratio between the magnetic field $h$ and the coupling constant $J_{\text{eff}}$.
\paragraph{The magnetically dominated regime:} This regime corresponds to $J_{\text{eff}}\in]-\infty,J_{\text{min}}]$ where $J_{\text{min}}$ is small compared to $h$. When minimizing the Hamiltonian \eqref{eq:Interpret Hamiltonian} the contribution of the magnetic field $h>0$ to the total Hamiltonian is minimized by maximizing $F$. For negative values of the effective coupling constant, i.e.~ $E>J$ the interaction term is also minimized by maximizing $V_\pa$. Thus the magnetic field and the interaction both maximise the size of the domain. We indeed obtain this spin configuration from the numerical simulation \ref{Fig:Maximal domain}. The minimal Hamiltonian becomes
\begin{equation}
   H_\text{min}=-hF_{\text{max}}+(J-E)V_{\pa,\text{max}}\,,
   \label{eq:GS_maximal}
\end{equation}
where $F_{\text{max}}$ is the size, and $V_{\pa,\text{max}}$ is the boundary length of the maximal domain. In this regime, the minimal Hamiltonian is linear in the effective coupling constant. This agrees with the functional behavior of the free energy at low temperature, which can be approximated by the minimal Hamiltonian (see Fig.~\ref{Fig:FreeEnergy}). We note that in this regime the numerical approach is only sufficient to match the qualitative behavior of the free energy. This is due to the fact that we necessarily have to consider lattices of finite size in the simulation. Furthermore, this regime does not include the original gravity theory, where the size of the spin domain has to be kept finite, as explained at the end of subsection \ref{subsec:Mapping}.  
\paragraph{The coupling dominated regime:} This regime corresponds to $J_{\text{eff}}\in[J_\text{max},\infty[$ where $J_\text{max}>J_\text{min}$ is a positive coupling constant that is large compared to $h$. For positive coupling constants, the interaction term of \eqref{eq:Interpret Hamiltonian} is minimized, by minimizing the size of the domain. Since the coupling constant is large compared to the magnetic field, which tries to maximize the size of the domain, the domain is minimized. We indeed observe this behavior in the numerical simulation Fig.~\ref{Fig:Minimal domain} and the minimal Hamiltonian reads
\begin{equation}
    H_{\text{min}}=0\,.
    \label{eq:GS_min}
\end{equation}
We again emphasize that the minimal Hamiltonian approximates the free energy at low temperature, which is constant for large couplings (see Fig.~\ref{Fig:FreeEnergy}). The constant shift between the numerical free energy, and the minimal Hamiltonian \eqref{eq:GS_min} is due to the fact that the numerical approach calculates the difference in free energy from a reference ensemble \eqref{eq:difference free energy}.  We note that this regime also does not describe the discrete gravity theory, where spin domains need to have finite size.

\paragraph{The intermediate regime:} This regime corresponds to small positive values of the effective coupling constants $J_\text{eff}\in]J_\text{min},J_\text{max}[$. For this regime, the contribution from the magnetic field still aims at minimizing the Hamiltonian by maximizing the size of the spin domain, while the contribution from the coupling aims at minimizing it. Thus, both effects balance each other, since the parameters are fine-tuned such that the Hamiltonian is minimized by a configuration in which the domain of spins pointing up is approaching the asymptotic boundary. Therefore, this gives rise to a plot similar to Fig.~\ref{Fig:Maximal domain}. As mentioned at the end of subsection \ref{subsec:Mapping}, to relate the original discrete JT gravity theory to the Ising model, the spin domain is not allowed to touch the asymptotic boundary. Consequently, the intermediate regime is identified as the relevant regime for describing the resolvent of discrete JT gravity. The resolvent is given in terms of the free energy as defined in \eqref{eq:Ising Free energy} with $E=J-J_\text{eff}$. In the plot of the free energy given in Fig.~\ref{Fig:FreeEnergy}), the intermediate regime corresponds to the part where the graph deviates from the asymptotic lines. We note that the system can only be in the ground state if the free energy at low temperature is lower than the energy corresponding to the asymptotic lines. The peak where the free energy is higher than the asymptotic lines is a numerical artifact, since when the magnetic field and the coupling constant balance each other out, the energy difference at each update step is almost zero. Thus, the time it takes to reach the actual ground state of the system diverges, and the numerical simulation breaks down.   A valid result for the resolvent of JT gravity is therefore obtained from the regime in which the free energy is lower than the asymptotic lines.

\section{Mean-field approximation from inflation}
\label{sec:mean field}
In this section, we consider the mean-field approximation of the classical Ising model on a network with foliation structure, such as the hyperbolic lattice generated by the inflation rule (see Fig.\ref{Fig 3,7 tiling}-\ref{Fig 7,3 tiling}). Each local Ising spin at each site has $p$ nearest neighbors. The general Ising Hamiltonian on this network is given by
\begin{align}\label{eq:MFH}
    H=-J\sum_{\avg{rr'}} s_rs_{r'}-\sum_{r} h_r s_r \, ,
\end{align}
For later convenience, we have considered a local magnetic field $h_r$ and written the Hamiltonian in a simplified form. We recover the original Hamiltonian in \eqref{eq:Hamiltonian Ising model} as $H_\eqref{eq:Hamiltonian Ising model}=  -\kc{\gamma_\phi+\frac q6} + \frac12\kc{J \mathcal U - h \mathcal N + H_{\eqref{eq:MFH}}|_{h_r=h}}$, where $\mathcal U=\sum_{\avg{rr^\prime}\in\Delta^\times}$ and $\mathcal N=\sum_{r\in\Delta^\times}$ are the numbers of links and sites in the hyperbolic lattice $\Delta^\times$ respectively.

\subsection{Foliating the mean field}
We label the foliation with index $n=1,2,\cdots,N+1$, where $n=1$ is the central layer and $n=N+1$ is the boundary layer, where we impose the condition $s_r=-1$ for the spins located at boundary. Layer $n$ has $l_n$ sites, where each site is linked to $z_n$ sites on the $n$-th layer itself, $u_n$ sites on the $(n+1)$-th layer, and $v_n$ sites on the $(n-1)$-th layer. We set $l_0=v_1=u_{N+1}=0$. These network parameters give the number of sites $\mathcal N=\sum_{n=1}^N l_n$ and the number of links $\mathcal U=\sum_{n=1}^N u_nl_n$. Furthermore they satisfy the constraints $l_nu_n=l_{n+1}v_{n+1}$ and $z_n+u_n+v_n=p$. We will only consider the layer-dependent magnetic field $h_n$.
We apply our mean-field approximation in two steps. In the first step, we neglect the fluctuation within each layer, such that all spins $s_r$ within the $n$-th layer are identical, denoted as $s_n$. This assumption may break down for $J<0$, where the system may enter the anti-ferromagnetic phase, subject to the boundary condition chosen. Thus we only consider $J\geq0$. We rewrite the Hamiltonian in terms of the spins on each layer as
\begin{equation}
    H_{\rm foliating}=-J \sum_{n=1}^N l_n s_n\kc{\frac12 z_n s_n+u_n s_{n+1}}-\sum_{n=1}^N h_n l_n s_n\,,
    \label{eq:foliating_Hamiltonian}
\end{equation}
where $s_{N+1}=-1$.
We define the magnetization density on the $n$-th layer as $\avg{s_n}=m_n$. In the mean-field approximation, we only consider small fluctuations around the local magnetization. Thus, we neglect the correlation between the fluctuations $(s_n-m_n)(s_{n'}-m_{n'})$ and obtain
\begin{align}
    s_ns_{n'}\approx s_n m_{n'}+m_n s_{n'} -m_n m_{n'}\,.
\end{align}
Inserting this condition into \eqref{eq:foliating_Hamiltonian} we find
\begin{equation}
    \begin{split}
        H_{MF}=&~-\frac12 J \kd{\sum_{n=1}^N l_n(2s_n-m_n)(z_n m_n+u_n m_{n+1}+v_n m_{n-1})
    +l_Nu_Nm_Nm_{N+1}}\\
    &~ -\sum_{n=1}^N h_nl_n s_n\,,
    \end{split}
\end{equation}
where $m_{N+1}=-1$. By summing over all possible spin configurations $\ke{s_n}$, we obtain the mean-field partition function as
\begin{align}
\label{eq:MFpartition}
    Z_{MF}
    =\sum_{\ke{s_n}_{n=1}^N}e^{-\beta H_{MF}}
    =&~\prod_{n=1}^N 2\cosh\ke{\beta l_n\kd{J(z_n m_n+u_n m_{n+1}+v_n m_{n-1})+h_n}} \\ 
&~~~~~\times\exp\ke{-\frac12 \beta J l_n m_n\kd{z_n m_n+u_n m_{n+1}(1-\delta_{nN})+v_n m_{n-1}}}\,. \nn
\end{align}
In the next subsection, we use this partition function to derive the self-consistency equation for the local magnetization in each layer, $\ke{m_n}$.
\subsection{Self-consistency equation}
The local magnetization for a uniform magnetic field $h_n=h$ is given by 
\begin{equation}
    \avg{s_n}=\frac{\sum_{\ke{s_n}_{n=1}^N}s_ne^{-\beta H_{MF}}}{\sum_{\ke{s_n}_{n=1}^N}e^{-\beta H_{MF}}}
    =\frac{1}{\beta l_n}\frac{\partial}{\partial h_n}\ln Z_{MF}\Big |_{h_n=h}
    \label{eq:local magnetization}
\end{equation}
for all $n=1,2,\cdots,N$. By inserting the mean-field partition function \eqref{eq:MFpartition} into \eqref{eq:local magnetization} we obtain the self-consistency equation
\begin{equation}\label{eq:saddle point}
    m_n=\avg{s_n}=\tanh \ke{\beta  l_n \kd{J (z_n m_n+u_n m_{n+1}+v_n m_{n-1})+h}}\,.
\end{equation}
We consider $h\geq 0$  for low temperatures, i.e.~$\beta\to \infty$, such that $\tanh(\beta x)\to \sgn(x)$. Recall the boundary condition $m_{N+1}=-1$. 
For solving the self-consistency equation, we consider the ansatz 
\begin{equation}\label{eq:MFAnsatz}
    m_n(w)=\begin{cases}
    1\,,& 1\leq n\leq w\,,\\
    -1\,,& w+1\leq n\leq N\,,\\
    \end{cases} \quad \text{for}\quad  
    w\in \mathbb Z,\quad 0\leq w \leq N,
\end{equation}
which has a domain wall between the $w$-th layer and the $(w+1)$-th layer. Inserting this into \eqref{eq:saddle point}, we obtain
\begin{equation}
    0=\begin{cases}
        \sgn(h+Jp)-1\,,& n\leq w-1\,,\\
        \sgn(h-Jp)+1\,,& n\geq w+2\,.\\
    \end{cases}
\end{equation}
There are two possible self-consistent solutions for which the magnetization assumes either the value  $m_n=1$ or $m_n=-1$. For both cases to be possible solutions, we need 
\begin{equation}
-Jp<h \quad \mathrm{and} \quad  h<Jp\,.
\label{eq:condition magnetic field}
\end{equation}
Otherwise, only one of the magnetizations is possible, set by the external magnetic field.
Restricting to positive $h$, under the above requirements, with $0\leq h<Jp$, the equations for the range $w\leq n\leq w+1$ are
\begin{equation}\label{eq:MFSaddle}
    0=\begin{cases}
        \sgn[h+J (p-2u_w)]-1,& n= w \, , \\
        \sgn[h-J (p-2v_{w+1})]+1,& n= w+1 \, .\\
    \end{cases}
\end{equation}
Therefore the ansatz \eqref{eq:MFAnsatz} solves \eqref{eq:saddle point} if
\begin{align}\label{eq:SaddleCondition}
    -J (p-2u_w)< h<J (p-2v_{w+1})\,.
\end{align} 
For a hyperbolic network, we have $v_n\leq u_n$ therefore $0\leq p-2v_n$ and $p-2u_n\leq p-2v_n$.
If $u_n$ and $v_n$ are constants, the ansatz $m_n(w)$ is a solution for any $w$ or none of the $w$. \\
We now consider values for $J,h$ outside of the range given by \eqref{eq:condition magnetic field}. For $0\leq Jp\leq h$, the magnetization $m_n=-1$ is not a self-consistent 
solution, which leads  to the spins pointing up $s_r=+1$ for $r=1,...,N$. However, the boundary condition enforces $s_{N+1}=-1$. Therefore, the ground state is given by all spins pointing up except for the spins located at the boundary. \\
For $J(p-2v_n)\leq h<Jp$ or $0<h\leq -J(p-2u_n)$, both $m_n=\pm1$ are self-consistent solutions $\forall n$. However, the saddle point equation \eqref{eq:MFSaddle} is not satisfied by these solutions. This means that in this case, the ansatz \eqref{eq:MFAnsatz} for a single domain wall located between two layers is too naive. it appears necessary to include fluctuations, allowing for a domain wall crossing different layers.\\
For $h\leq -Jq $ and $ 0\leq J$, the case that all spins point up, i.e.~$m_n=1$ is not a self-consistent solution. Therefore, the ansatz $m_n(0)$ is the only solution, corresponding to the configuration of $m_n=-1, \ \forall n$. Actually, for $h\leq 0< J$,  the ground state is also $m_n(0)$ due to the magnetic field and the boundary condition.

\subsection{Energy analysis}
Assume that the condition \eqref{eq:SaddleCondition} holds for any $w$, i.e.~the ansatz \eqref{eq:MFAnsatz} for any $w$ is an on-shell solution. The on-shell energy corresponding to $m_n(w)$ is
\begin{equation}\label{eq:MFEnergy}
\begin{split}
    H(w)=&~ -J \sum_{n=1}^N l_n m_n\kc{\frac12 z_n m_n+u_n m_{n+1}}-h\sum_{n=1}^N l_n m_n\\
    =&~ -\kc{\frac12 Jp-h}\sum_{n=1}^N l_n 
    -\frac12 J\kc{l_N u_N-l_1v_1}- 2h \sum_{n=1}^w l_n 
    +2J u_{w}l_{w}\,.
\end{split}
\end{equation}
The energy difference between two different solutions of the saddle point equation $m_n(w)$ and $m_n(w-1)$ is given by
\begin{equation}\label{eq:MFEnergyDifference}
\begin{split}
    H(w)-H(w-1)
    =&~ 2(J(u_{w}-v_{w})-h)l_{w}\,.
\end{split}
\end{equation}
If $u_n$ and $v_n$ are constants, depending on the sign of $(J(u_{w}-v_{w})-h)$, the ground state is given by all spins pointing either up or down.

\subsection{Foliation from inflation rule} 

For a fixed \pq{q}{p} lattice, the number of sides connecting to the previous and next layer can be derived as follows. We only consider the network where each spin has $p$ neighboring spins connected by links. 
As explained in section \ref{subsec:inflationTiling},
the inflation-generated network  has a natural foliation for which the coefficients $u_n,v_n$ are determined by the inflation rule.\\
Based on the connection to the previous layer, we consider $\alpha$ types of spins $\{a_i|i=1,2,\cdots,\alpha\}$, where the $a_i$-type spin is connected to $x_i$ number of spins in the previous layer. For each type of spins $a_i$ we impose an inflation rule $a_i\to w_i$, where the word $w_i$ is a sequence of letters $\ke{a_j}$. A word $w_i$ contains $k_i$ number of general letters and $k_{ij}$ number of letters $a_j$. We have $k_i=\sum_{j} k_{ij}$ such that $M=\ke{k_{ij}}_{ji}$ is the  substitution matrix, which for a \pq{q}{3} lattice is given by \eqref{eq:InflationMatrix}.\\
Consider the $n$-th layer as a closed sequence $\sigma_n$ with $l_{ni}$ number of $a_i$. The average connection coefficients are
\begin{align}
    u_n=\frac{\sum_{ij}l_{ni}k_{ij}x_j}{\sum_{i}l_{ni}}\,,\quad
    v_n=\frac{\sum_{i}l_{ni}x_r}{\sum_{i}l_{ni}}\,,
\end{align}
The number of spins of neighboring layers is related by
\begin{align}
    l_{n+1,i}=\sum_{j}l_{nj}k_{ji}\,.
\end{align}
They automatically satisfy the constraint $l_n u_n=l_{n+1}v_{n+1}$.

\subsection{Comparison with the Monte Carlo approach}
\begin{figure}
    \centering
    \includegraphics[width=0.65\linewidth]{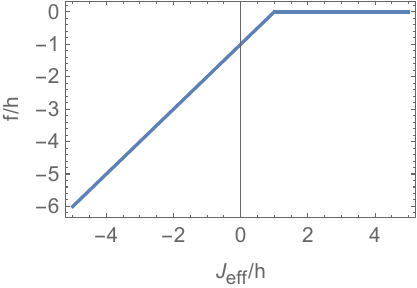}
    \caption{The shifted free energy density $f=(H_{\eqref{eq:Hamiltonian Ising model}}+(\gamma_\phi+q/6))/\mathcal{N}$ as a function of the effective coupling $J_{\rm eff}$ in units of the magnetic field $h$ from the mean field approximation on the tree network. For $J_\text{eff}>h$ the ground state of the system is given by all spins pointing down, and the free energy is constant. In the case of $J_\text{eff}<h$ the ground state corresponds to all spins pointing up, while the free energy is linear. The turning point, i.e.~the intermediate regime, is located at $J_{\text{eff} } / h = 1$.}
    \label{fig:MFEnergy}
\end{figure}
For simplicity, we focus on the mean-field approximation of the hyperbolic lattice at large $q$, which is the number of vertices in every polygon.
The foliation of the mean-field approximation is not necessarily the same as the foliation obtained from the inflation rule in the previous subsection. 
We note as a crucial point that for large $q$, 
the mean-field foliation on this network is approximated by the foliation of a tree network with  $\alpha=1,\ k_{11}=2,\ x_1=1,\, l_n=2^{n-1}$, as well as $p=3,\, z_n=0,\, u_n=2,\ v_n=1$. 
The energy difference in \eqref{eq:MFEnergyDifference} becomes 
\begin{equation}
    H(w)-H(w-1)=2(J-h)2^w\,.
\end{equation}
Therefore the ground state is given by all spins pointing down when $J>h$ and all spins pointing up when $J<h$. Thus, the transition point is located at $J=h$. The ground state energy \eqref{eq:MFEnergy} is given by
\begin{equation}
    H_{\eqref{eq:MFEnergy}}=\begin{cases}
        \left(2^m-1\right) (h-2 J), & J\geq h\,, \\
        h \left(1-2^m\right)+2 J, & J<h\,. \\
    \end{cases}
\end{equation}
As noted in the beginning of this section, we use different conventions for the mean-field Hamiltonian \eqref{eq:MFH} and the Hamiltonian of the Ising model \eqref{eq:Hamiltonian Ising model} in the previous section. The later is given in terms by the former as $H_\eqref{eq:Hamiltonian Ising model}=  -\kc{\gamma_\phi+\frac q6} + \frac12\kc{J \mathcal U - h \mathcal N + H_{\eqref{eq:MFH}}|_{h_r=h}}$, where $\mathcal U=\sum_{\avg{rr^\prime}\in\Delta^\times}$ and $\mathcal N=\sum_{r\in\Delta^\times}$ are the numbers of links and sites in the hyperbolic lattice $\Delta^\times$ respectively. We thus obtain the minimal Hamiltonian of the Ising model \eqref{eq:Hamiltonian Ising model} from the mean-field result \eqref{eq:MFEnergy} as
\begin{equation}
    H_{\eqref{eq:Hamiltonian Ising model}}+\kc{\gamma_\phi+\frac q6}
    =\begin{cases}
        0,& J_{\rm eff}\geq h \, , \\
        h+(J_{\rm eff}-h) 2^m, & J_{\rm eff}<h \, ,
    \end{cases}
\end{equation}
where we shifted the Hamiltonian by $\gamma_\phi+q/6$ to obtain a simpler expression. In the low-temperature limit, the free energy is equivalent to the ground state energy, as can be seen from \eqref{eq:Free energy, internal energy}. The mean-field approximation of the free energy density for large \( m \) is illustrated in Fig.~\ref{fig:MFEnergy}. Its asymptotic behavior qualitatively matches the numerical results shown in Fig.~\ref{Fig:FreeEnergy} on both sides. Moreover, since the Monte Carlo approach takes fluctuations into account the free energy is lower than in the mean-field approach in the asymptotic regime. Near the turning point \( J_{\rm eff} = h \), the behavior of the free energy differs between both approaches. This discrepancy may be due to the breakdown of our mean-field approximation and the chosen ansatz. As the energy \( H(w) \) becomes less sensitive to the position of the domain wall around the turning point, the domain wall can bend with minimal energy cost and thus easily deviate from the foliated ansatz \eqref{eq:MFAnsatz}. These deviations are observed in the Monte Carlo simulation, and they lead to a state that is given by a single domain of spins pointing up, which has finite size. The energy of this state is lower than the energy of the state predicted by the mean-field approximation. Additionally, our mean-field model only considers a tree network, which corresponds to the \( q \to \infty \) case. Consequently, the results will deviate from those at finite \( q \), particularly when the size of the domain wall is comparable to the size of a \( q \)-sided polygon in the network. Another reason for the discrepancy between both approaches is that the numerical approach is no longer applicable in the regime where the free energy has its peak, as previously explained. To summarize the comparison of Monte Carlo and mean-field approaches, we emphasize that the predictions of both approaches agree in the asymptotic regimes, while for the intermediate regime only the numerical approach gives results that agree with the gravity expectation.

\section{Discussion and outlook}
\label{Sec:Conclussion}
In this paper, we present a further step towards a discrete holographic duality involving a dynamical gravity theory on a hyperbolic lattice. We consider a discrete analog of JT gravity \eqref{eq:Discrete JT action}, recalling that in the continuous case, the holographic dual of this model is known to be a matrix model \cite{Saad2019}. Similar to the continuous case, the topological expansion of the path integral in the discrete case is dominated by the disk topology. The disk partition function is given by a sum over closed curves in a hyperbolic lattice, with each path being weighted by the enclosed area \eqref{eq:Calculation Disk}. We map this sum over closed paths to an equivalent sum over spin configurations on the dual lattice \eqref{eq:result discrete disk partition}, where the spins pointing up have to form a single simply connected domain \eqref{eq:Constraint disk topology}. This leads to a hyperbolic Ising model \eqref{eq:Hamiltonian Ising model}, with its partition function determined by the resolvent of discrete JT gravity \eqref{eq:Resolvent and partition function}. The resolvent is obtained from the disk partition function via a Laplace transformation, where we integrate all possible values of the asymptotic boundary length. Due to its origin from the disk path integral, the allowed spin configurations of the Ising model are restricted through the topological constraint, which makes this model very different from the ordinary hyperbolic Ising model \cite{breuckmann2020critical,Asaduzzaman:2021bcw}. Furthermore, the parameters are determined by the gravity theory, and the low-temperature regime of the Ising model corresponds to the semiclassical limit \eqref{eq:Magnetic field and coupling constant}.\\
We study the low-temperature regime of the Ising model numerically. This necessarily introduces a radial cutoff to the hyperbolic lattice. As emphasized throughout this manuscript, at finite truncation of the lattice the relation between discrete JT gravity and the Ising model only holds when the domain wall between spins pointing in opposite directions is located away from the asymptotic boundary. This regime is determined by the numerical analysis on a finite truncation. We compute the free energy \eqref{eq:Ising Free energy} using a Monte Carlo method at low temperatures. The free energy as a function of the coupling constant is shown in Fig.~\ref{Fig:FreeEnergy}. We identify three phases in the Ising model's phase structure, governed by the ratio of the coupling constant and external magnetic field. In the extreme regimes, characterized by a large positive or small positive/negative coupling constant, the ground state of the system corresponds to a spin domain of maximal or minimal size, respectively, as can be seen in Fig.~\ref{Fig:Maximal domain}-\ref{Fig:Minimal domain}. In the intermediate regime, the spin domain possesses a finite size. This can only correspond to the ground state if the free energy is lower than the energy of the asymptotic lines, which is indeed satisfied for certain values of the coupling constant. For those values for which the functional behavior of the free energy exceeds the asymptotic line, the system is not in the ground state. As emphasized throughout the paper, discrete JT gravity corresponds to the case that the ground state of the system is given by a single domain of spins pointing up, which has finite size. The functional behavior of the free energy in this regime determines the resolvent function of the disk. Moreover, these numerical results are validated through a mean-field approximation, which is based on the foliation structure of the underlying lattice. The free energy density obtained from this approach is shown in Fig.~\ref{fig:MFEnergy}, and agrees with the numerical predictions in the asymptotic regime. As discussed at the end of section \ref{sec:mean field}, the mean-field approach does not capture the intermediate regime.\\ 
Truncating the infinite hyperbolic lattice to a finite one bears similarities to the approach of  \cite{Stanford2020,Iliesiu:2020zld}, where continuous JT gravity is studied at finite cutoff. In \cite{Stanford2020}, the evaluation of the disk partition function is mapped to a stochastical problem, involving self-avoiding random walks. It is instructive to compare that approach to ours. As explained in section \ref{sec:Discrete JT} above, both for the discrete and the continuous case the disk partition function is obtained from a path integral performed over all possible closed paths, as illustrated in Fig.~\ref{Fig boundary wiggles}. In \cite{Stanford2020}, each path is interpreted as a closed random walk. To impose the disk topology, the random walk has to be self-avoiding. In our case, the sum over closed boundary paths is interpreted as a sum over spin configurations, as illustrated in Fig.~\ref{Fig:Map to Ising}, and the disk topology is imposed by the constraint that the spins pointing up have to form a single simply-connected domain. 
In \cite{Stanford2020} the disk partition function is evaluated in the semiclassical limit, which corresponds to small $G$ and a random walk that is close to the asymptotic boundary. Depending on the size of microscopic fluctuations of the walk, three different regimes are identified there as well, albeit with a different nature than ours. These are the Schwarzian regime, the flat space regime, and an intermediate regime, that arise from large, small, and intermediate loop sizes, respectively. In our work, we also consider the semiclassical limit, which from the point of the Ising model corresponds to a low-temperature limit. While we also find three different regimes in the phase structure of the Ising model, these are very different from the ones found in \cite{Stanford2020}. In particular, the two extremal regimes in our work in which the domain wall shrinks to zero size or touches the asymptotic boundary are not present in \cite{Stanford2020}. This is due to the fact that in that work only random walks of macroscopic size are considered. On the other hand, our work cannot reproduce the the flat space and intermediate regimes of \cite{Stanford2020} that correspond to microscopic fluctuations of the boundary. This is due to the fact that the Ising model has a finite lattice spacing, and thus the size of fluctuations is limited by the the edge length of the triangles. \\

We conclude this work by commenting on possible extensions of our analysis, which have to be investigated in order to establish a full holographic duality for discrete spacetime.\\
The most promising avenue in this direction is to test whether the known duality between continuous JT gravity and a matrix model can still be realized in the discrete case \cite{Saad2019}. This duality arises by showing that the genus expansion of the correlation function $\langle Z(\beta_1)...Z(\beta_n)\rangle$ coincides with the topological expansion of the corresponding $n$-point function of the matrix model. The expectation value $\langle...\rangle$ is taken with respect to the gravity path integral. The $g$-th order of this expansion is the partition functions of JT gravity on a hyperbolic Riemann surface with $n$ boundaries of length $\beta_1,...,\beta_n$ and genus $g$. These partition functions are obtained from the Weil-Petersson volumes of the underlying Riemann surface, which satisfy  Mirzakhani’s recursion relation \cite{Mirzakhani2007}. On the other hand, the perturbative expansion of matrix ensemble correlation functions is governed by so-called ''loop equations'' \cite{MIGDAL1983199}. The loop equations also satisfy a recursion relation \cite{Eynard2004}, which is a particular example of a topological recursion \cite{EynardOranti2007}. The leading eigenvalue density $\rho_0(E)$, of the matrix model serves as input data of the recursion relation, which thus fixes the correlation functions to all orders in perturbation theory. A key fact for establishing the JT/matrix model duality is that the Mirzhakani recursion relation is related to the topological recursion relation by a Laplace transformation \cite{Eynard2007}. In particular, $\rho_0(E)$ is determined from the disk partition function $Z_{\text{disk}}(\beta)$, which is the leading order of the genus expansion of $\langle Z(\beta)\rangle$. Thus it is sufficient to know the disk partition function of JT gravity in order to determine the holographic dual. Unlike holographic dualities in higher dimensions, the JT/matrix model duality can be explicitly proven at the perturbative level, since all correlation functions match to all orders in perturbation theory.\\
Since the leading eigenvalue density is related to the resolvent by an integral transformation, namely a Stieltjes transformation, we may extract it from the Ising free energy, and propose that discrete JT gravity is dual to a matrix model with this spectral density. In order to test this proposed duality, we will have to compute the resolvent function of discrete JT gravity on general hyperbolic Riemann surfaces, and check if the results agree with the matrix model prediction. In this context, it will be crucial to compute the wormhole partition function, since its dual quantity is a universal expression that does not depend on the concrete matrix model \cite{Saad2019}. The wormhole geometry corresponds to a Cauchy slice of the BTZ black hole, passing through the bifurcation point \cite{PhysRevLett.69.1849,PhysRevD.51.622}. The analogs of these wormhole geometries for a hyperbolic lattice have recently been studied in the literature \cite{Chen:2023cad,Dey:2024jno}. In our case, the wormhole partition function may be computed by considering the corresponding Ising model on such a lattice.\\
A further approach worth pursueing is to relate discrete JT gravity to aperiodic spin chains, which were studied as boundary theories for discrete holography \cite{Basteiro2022,Basteiro2022a}. There,  the entanglement entropy for these spin chains was computed using an SDRG (strong disorder RG) approach. Since the spin chain considered is weakly coupled, the spin chain result for the central charge in the entanglement entropy disagrees with the bulk result obtained from a discrete analog of the Ryu-Takayanagi formula \cite{Ryu:2006bv,Ryu:2006ef}. In this context it would be interesting to consider corrections from bulk entanglement, in analogy to \cite{Faulkner:2013ana}. For discrete holography, these corrections may be calculated from the Ising model description of discrete JT gravity proposed in this paper. A promising candidate for a strongly-coupled boundary theory is given by an aperiodic SYK model \cite{SachdevYe:93,Kitaev:2015a}. Due to the known relation between continuum JT gravity and the SYK model \cite{Kitaev:2015a,Polchinski2016,Maldacena2016}, it will be interesting to investigate their relation in the discrete case. \\
A further idea is to compare our approach to discrete holography with \textbf{p}-adic AdS/CFT \cite{Gubser:2016guj,Heydeman:2016ldy,Heydeman:2018qty,Hung:2019zsk}. In this context, a Cauchy slice of the discrete bulk geometry is the so-called \textit{Bruhat-Tits tree}, which is dual to a theory defined on the \textbf{p}-adic numbers $\mathds{Q}_{\textbf{p}}$, with \textbf{p} prime. A Bruhat-Tits tree can formally be identified with the $p\to\infty$ limit of a \pq{p}{q} tiling, where $q=\textbf{p}+1$ \cite{Basteiro2022}. A dynamical gravity theory can be defined on the Bruhat-Tits tree by considering edge length fluctuations of the graph \cite{Gubser2016}. Since these trees correspond to the spatial slice of a three-dimensional geometry, we have to extend our approach to include the temporal direction in order to compare both approaches.\\
Finally, we comment on possible extensions of our bulk theory to higher dimensions. Since the hyperbolic plane may be viewed as a time slice of AdS$_3$, we may include the Lorentzian time direction through a construction in the spirit of causal dynamical triangulation \cite{Ambjorn:2012jv,Loll:2019rdj}. In this context, we have to consider a stack of several \pq{3}{q} lattices, where different lattices are connected through timelike edges. A different way is  to start from pure gravity defined in Euclidean AdS$_3$ and perform a dimensional reduction. This leads to an Einstein-Maxwell-dilaton theory in two dimensions \cite{Maxfield2020}. In the discrete case, we thus have to add a gauge field to the action, and we can study the properties of the higher-dimensional theory through this approach. We may in particular consider the matrix model dual to the discrete Einstein-Maxwell-dilaton theory, which is known in the continuous case \cite{Maxfield2020}. We leave the investigation of these open questions for future research.\\

\acknowledgments

We are grateful to Pablo Basteiro, Giuseppe Di Giulio, Fabian Köhler, Ren\'e Meyer, Roderich Moessner and Zhenbin Yang for useful discussions. We acknowledge support by the Deutsche Forschungsgemeinschaft (DFG, German Research Foundation) under Germany's Excellence Strategy through the W\"urzburg-Dresden Cluster of Excellence on Complexity and Topology in Quantum Matter - ct.qmat (EXC 2147, project-id 390858490). The work of   J.E.~and Z-Y.X~is also supported by the Collaborative Research Centre SFB 1170 `ToCoTronics', project-id 258499086. The work of J.~E.~and Y.~T.~is also supported by DFG grant ER 301/8-1  and by the DIP grant `Holography and the Swampland.'
Z-Y.X.~also acknowledges support from the National Natural Science Foundation of China under Grant No.~12075298.

\appendix

\section{Counterterm from inflation rules}
\label{apx:counterterm}

When considering the Schwarzian regime of JT gravity we have to take a limit of large boundary length $\mathcal{L}\to\infty$. Since the Gibbons Hawking term in \eqref{eq:conti JT} diverges for this limit we have to subtract a counterterm to ensure a well defined large distance behavior. The same holds true in the discrete case, since the boundary term in \eqref{eq:Discrete JT action} diverges, for large lattices, i.e.~for a large number of boundary vertices $V_\pa\gg1$. Since we consider JT gravity on a \pq{3}{q} lattice the asymptotic behaviour of the boundary action is determined by the largest eigenvalue, and the right eigenvector of the inflation matrix \eqref{eq:InflationMatrix}. We have
\begin{equation}
    S_{\text{bdr}}\sim\sum\limits_{v\in\pa\Delta}\psi_v=V_{\pa,a}\psi_a+V_{\pa,b}\psi_b=(\psi_ap_a+\psi_bp_b)V_\pa\,,
    \label{eq:Boundary action}
\end{equation}
where $\psi_{a,b}$ is the deficit angle associated to vertices with letters $a,b$, and $V_{\pa,a/b}$ are the number of boundary vertices with the corresponding letter. In the second equality we used that the components of the right eigenvector of the inflation matrix \eqref{eq:InflationMatrix}, namely $p_{a,b}$ encode the frequency of letters in the asymptotic series. In case of a \pq{3}{q} lattice we have 
\begin{equation}
    \begin{split}
         (\psi_a,\psi_b)&=(\pi,\pi)-(2\pi/3,\pi)\,,\\
         (p_a,p_b)&=\frac{1}{2}\left(\sqrt{(q-2)(q-6)}-q+6\,,-\sqrt{(q-2)(q-6)}+q-4\right)\,.
    \end{split}
\end{equation}
The divergence of the boundary action for $V_\pa\to\infty$ is cancelled by the introduction of the counterterm

\begin{equation}
   \psi_c=(\psi_a,\psi_b)\cdot(p_a,p_b)=\frac{\pi}{6} \kc{-q+\sqrt{(q-6) (q-2)}+6}\,. 
   \label{eq:counterterm}
\end{equation}

\bibliographystyle{JHEP}
\bibliography{bib.bib}

\end{document}